\documentclass[preprints,article,accept,pdftex,moreauthors]{Definitions/mdpi} 

\usepackage{enumitem}

\firstpage{1} 
\pubvolume{1}
\issuenum{1}
\articlenumber{0}
\pubyear{2026}
\copyrightyear{2026}
\externaleditor{Name} 
\datereceived{7 June 2026} 
\daterevised{6 July 2026} 
\dateaccepted{9 July 2026} 
\datepublished{ } 
\pdfoutput=1 

\usepackage{longtable}
\usepackage{mathcomp}
\usepackage{lipsum}
\usepackage{placeins}
\usepackage{textcomp}
\usepackage{latexsym}
\usepackage{gensymb}
\newcommand*\aap{Astron. Astrophys.}

\newcommand*\aaps{Astron. Astrophys. Suppl. Ser.}

\newcommand*\aj{Astron. J.}

\newcommand*\apj{Astrophys. J.}
\newcommand*\apjl{Astrophys. J. Lett.}

\newcommand*\apjs{Astrophys. J. Suppl. Ser.}

\newcommand*\fcp{Fund.~Cosmic~Phys.}

\newcommand*\mnras{Mon. Not. R. Astron. Soc.}

\newcommand*\nat{Nature}

\newcommand*\pasj{Ublications Astron. Soc. Jpn.}
\newcommand*\pasp{Publ. Astron. Soc. Pac.}
\newcommand*\physrep{Phys.~Rep.}

\Title{Tidal Structures Around Edge-On Galaxies in \linebreak Deep Imaging Surveys}

\Author{{Kyle R. Adams} $^{1,}$*\orcidA{}, Aleksandr Mosenkov $^{1,}$*\orcidB{}, Jonah Seguine $^{1}$ \orcidC{}, Lydia Stacey $^{1}$, Thea \linebreak Spigarelli $^{1}$ and Jonah George $^{2}$\orcidD{}}

\AuthorNames{Kyle R. Adams, Aleksandr Mosenkov, Jonah Seguine, Lydia Stacey, Thea Spigarelli, and Jonah George}

\address{%
$^{1}$ \quad Department of Physics and Astronomy, Brigham Young University, Provo, UT 84602, USA; jseguine@student.byu.edu (J.S.); lstacey1@student.byu.edu (L.S.); theaes@student.byu.edu (T.S.) \\

$^{2}$ \quad Department of Astronomy, University of Maryland, College Park, MD 20742, USA; jonahg17@umd.edu\\} 
\corres{Correspondence: kyle012a@gmail.com (K.R.A.);  aleksandr\_mosenkov@byu.edu (A.M.)}

\abstract{
We present a statistical study of low-surface-brightness (LSB) tidal structures in two large samples of edge-on disk galaxies. Our primary sample comprises 5606 galaxies from the {Edge-on Galaxies In SDSS (EGIS)} catalog, analyzed using imaging from the DESI Legacy Imaging Surveys, supplemented by Hyper Suprime-Cam Subaru Strategic Program (HSC-SSP) data and deep Apache Point Observatory {(APO)} follow-up observations for selected objects. To assess the robustness of our results, we also examine an independent sample of 14{,}237 galaxies from the {Edge-on Galaxies in the Pan-STARRS survey (EGIPS)} catalog. All images were processed using a homogeneous procedure optimized for the detection of faint diffuse emission. Tidal structures were identified through visual inspection and classified into established morphological categories, with careful treatment of imaging artifacts and galactic cirrus contamination. We detected tidal features in 324~EGIS galaxies and 690 EGIPS galaxies, corresponding to incidence rates of 5.8\% and 4.8\%, respectively. Restricting the analysis to completeness-limited subsamples yields consistent fractions of 6.4\% and 6.2\%. At a typical DESI $r$-band surface-brightness depth of 28.6~mag\,arcsec$^{-2}$ {these} values are consistent with previous observational studies but lower than predictions from many cosmological simulations. Recent high-resolution simulations, however, produce incidence rates much closer to those measured here, suggesting that numerical resolution, realistic modeling of observational and instrumental effects, and galaxy formation physics are all critical for accurately predicting the abundance of LSB tidal structures.
}


\begin{document}


\section{Introduction}
\label{sec:intro}

Over the past decade, a significant advance has occurred in our ability to detect faint structures surrounding low-redshift galaxies. Modest-aperture telescopes equipped with wide fields of view~\citep{2010AJ....140..962M,2014PASP..126...55A,2019MNRAS.490.1539R}, complemented by targeted observations from larger \mbox{facilities~\citep{2016ApJ...823..123T,2021A&A...654A..40T,2025MNRAS.541.3015S},} now employ specialized observational and data-reduction techniques {(see, e.g., } 
\citep{2017ASSL..434..255K,2019arXiv190909456M}) to achieve the sensitivity required for detecting extremely low surface brightness (LSB) features. These approaches routinely reach surface brightness limits of $\sim$30~mag\,arcsec$^{-2}$~\citep{2019MNRAS.490.1539R}, and in some cases probe even fainter regimes~\citep{2014ApJ...782L..24V,2019arXiv191111579S,2021A&A...654A..40T}.

In parallel, deep imaging surveys have become essential tools for exploring the LSB Universe (see, e.g., \citep{2016MNRAS.456.1359F,2023A&A...671A.141M,2026MNRAS.546ag203D}). Large-area programs such as the DESI Legacy Imaging Surveys (DESI; Dey et al. \citep{2019AJ....157..168D}) and the Hyper Suprime-Cam Subaru Strategic Program (HSC-SSP; Aihara et al. \citep{2022PASJ...74..247A}) have dramatically expanded the volume and depth of available extragalactic data, enabling systematic searches for diffuse and previously undetectable structures in the outskirts of galaxies. Looking ahead, the Vera C.\ Rubin Observatory Legacy Survey of Space and Time (LSST, Ivezić et al. \citep{2019ApJ...873..111I}) is expected to further transform this field by delivering unprecedented depth, sky coverage, and temporal sampling, thereby providing a uniquely powerful dataset for studying LSB phenomena on a statistical basis~\citep{Martin2022,2025RNAAS...9..292R,2026RNAAS..10....5J}.

The rapid improvement in sensitivity to diffuse emission has important implications for our understanding of galaxy formation and evolution. Cosmological hydrodynamical simulations predict that galaxy interactions produce characteristic LSB structures, such as tidal streams and tails, shells, and stellar halos~\citep{1972ApJ...178..623T,Quinn1984,2019A&A...632A.122M}. Observational studies have increasingly confirmed these predictions using deep imaging data (e.g., \citep{2000A&AS..144...85S,2010AJ....140..962M,Jackson2023,2025A&A...700A.176M}), while further support is provided by detailed comparisons with numerical simulations\mbox{ (e.g., \citep{2015MNRAS.446..521S,2019MNRAS.490.3196P,2019A&A...632A.122M}). }
These LSB features act as fossil records of past galactic interactions, enabling the reconstruction of the assembly histories of galaxies. Their characterization therefore provides critical constraints on the dominant physical processes driving galaxy evolution across cosmic time and offers a direct means of testing theoretical models of structure formation in the Universe~\citep{Martin2022}.

Within this context, edge-on galaxies represent particularly advantageous targets for the detection and characterization of LSB structures in galactic outskirts~\citep{2002AJ....124.1328D,2002MNRAS.334..646K,2010MNRAS.401..559M,2014ApJ...787...24B}. {In addition to enabling measurements of disk thickness~\citep{2025MNRAS.540.3493T,2026ApJS..283...35Y}, their orientation also enhances} the visibility of faint features extending above and below the galactic mid-plane, which are often difficult to identify in face-on galaxies where such structures are projected onto the disk and diluted by the brighter central regions~\citep{2019ApJ...883L..32V,2020MNRAS.494.1751M,2020ApJ...897..108G,2020MNRAS.497.2039M,2021MNRAS.506.5030M,2022ApJ...932...44G,2022MNRAS.515.5698M}. 
In addition, edge-on galaxies reduce uncertainties associated with inclination effects, enabling more robust structural classifications~\citep{2020MNRAS.494.1751M}. The clearer separation between planar and extraplanar components in these galaxies facilitates detailed morphological and dynamical analyses of LSB features. As a result, edge-on galaxies provide a powerful laboratory for probing the signatures of past interactions and for constraining the physical processes that drive galaxy assembly and evolution.

Deep imaging surveys have also revealed a variety of faint foreground structures, most notably galactic cirrus clouds, which are of particular concern due to their diffuse nature and filamentary morphology, which can closely mimic genuine extragalactic LSB features~\citep{2010A&A...516A..83S}. This morphological similarity makes cirrus emission a significant source of contamination in deep imaging data, complicating the reliable identification and interpretation of faint structures in the outskirts of galaxies (e.g., \citep{2016A&A...593A...4M,2020A&A...644A..42R,2021MNRAS.508.5825M,2023MNRAS.519.4735S,2025A&A...704A.269L,2026A&C....5501075P}). 
Accounting for the presence of galactic cirrus is therefore essential when analyzing LSB features, particularly in studies aimed at identifying tidal debris and reconstructing the interaction histories of~galaxies.

The primary objective of this study is to quantify the occurrence rate and characterize the properties of LSB features in edge-on galaxies using the largest available {catalogs} ({{it should be noted that the} 
 Edge-on Galaxies in the DESI survey (EGIDE; \url{https://arxiv.org/abs/2606.16734} {accessed on 6 July 2026)} 
 was released during the review of this manuscript and now exceeds EGIS/EGIPS in size; this work uses the original {catalogs}
}) of bona fide edge-on disk galaxies. To this end, we utilize deep imaging from DESI and the HSC-SSP, enabling robust and consistent identification of faint tidal structures. We further compare the inferred occurrence fractions with results from previous observational studies as well as predictions from cosmological simulations, in order to assess the role of galaxy interactions in shaping the low-redshift Universe. A pilot study of LSB tidal features in edge-on galaxies was presented by \citet{Stripe82Paper}, based on a smaller sample drawn from the SDSS {(Sloan Digital Sky Survey)} 
Stripe~82 survey~\citep{2016MNRAS.456.1359F}. The present work builds upon that study by adopting a broadly similar methodology while extending the analysis to a substantially larger galaxy sample.
 
The remainder of this paper is organized as follows. In Section~\ref{sec:sample}, we describe the disk galaxy samples used in this study. The observational data are presented in Section~\ref{sec:data}, and the data preparation procedures are detailed in Section~\ref{sec:imgprep}. Our classification of LSB structures is given in Section~\ref{sec:classification}. The results are compared with previous studies and discussed in Section~\ref{sec:discussion}. Finally, our conclusions are summarized in Section~\ref{sec:conclusions}.

\section{The Samples}
\label{sec:sample}

Our main sample for LSB identification consists of 5{,}745 true edge-on galaxies drawn from the EGIS catalog (Edge-On Galaxies in SDSS; \citet{2014ApJ...787...24B}). The EGIS catalog was constructed from SDSS DR7~\citep{2009ApJS..182..543A} through visual inspection, yielding 4768 bona fide edge-on galaxies, and later expanded to 5747 objects by incorporating objects from complementary catalogs within the SDSS footprint~\citep{1999BSAO...47....5K,2011A&A...532A..74B,1991rc3..book.....D,2011MNRAS.410..166L}. \citet{2014ApJ...787...24B} showed that the catalog is $\sim$$95\%$ complete for galaxies with major-axis diameters $\gtrsim28$~arcsec, based on the $V/V_m$ test~\citep{1979ApJ...231..680T}. 

As a larger and more recent dataset of edge-on galaxies became available, we extended our analysis by incorporating the EGIPS catalog of 16{,}551 edge-on galaxies identified in the {Panoramic Survey Telescope and Rapid Response System (Pan-STARRS)} 
survey~\citep{2022MNRAS.511.3063M}. This catalog was constructed using a convolutional neural network trained on EGIS and HyperLeda data~\citep{2014A&A...570A..13M}, followed by visual inspection to ensure a reliable selection of genuine edge-on galaxies. The resulting sample is estimated to be $\sim$$96\%$ complete for galaxies with an $r$-band semi-major axis $> 6$~arcsec.

The EGIS and EGIPS catalogs share 3211 galaxies in common, as they are constructed from SDSS and Pan-STARRS surveys that cover slightly different, though significantly overlapping, regions of the sky. In this sense, despite their substantial overlap, the two catalogs are complementary.

In this work, we primarily base our analysis on the EGIS catalog, as the study was initially designed around this dataset. However, we also compare the global statistics of LSB features between the EGIS and EGIPS catalogs to support and validate our conclusions.

\section{The Data}
\label{sec:data}

Following the approach of \citet{Stripe82Paper}, who combined multiple imaging surveys for comparison, we utilize data from DESI DR9 and DR10~\citep{2019AJ....157..168D} and HSC-SSP PDR2 and PDR3~\citep{2022PASJ...74..247A}. This combination improves our classification of tidal and other structural features by incorporating deeper and higher-resolution imaging.

\begin{figure}[H]
\isPreprints{\centering}{} 
\includegraphics[width=11cm]{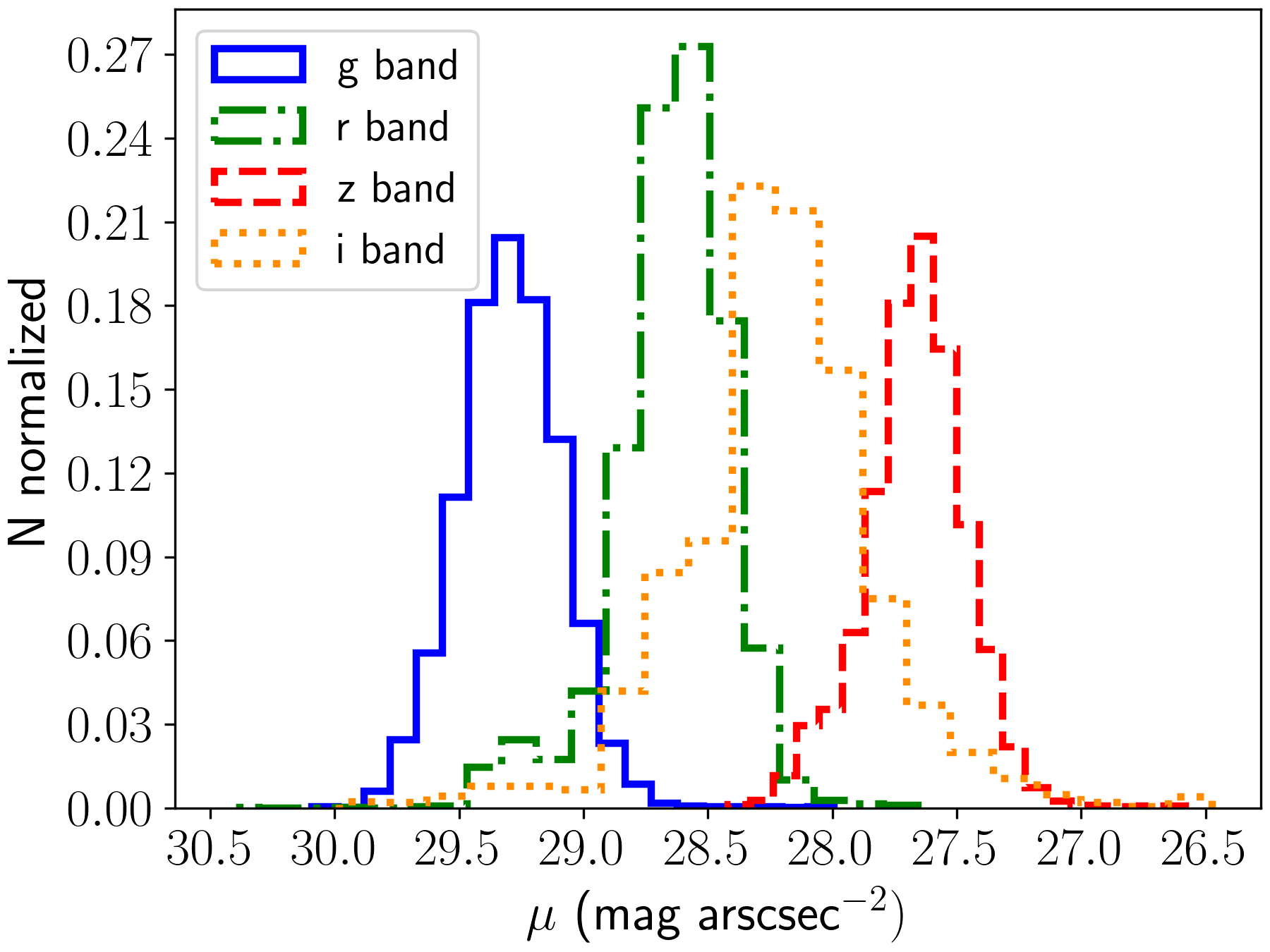}\\
\caption{Distribution of photometric depth (surface brightness) across all wavebands of the DESI data for the sample galaxies, measured within a $10''\times10''$ box at the $3\sigma$ level.}
\label{fig:desi_sb_egis}
\end{figure}

The DESI Legacy Imaging Surveys DR10, comprising DECaLS, BASS, and MzLS, cover more than 20{,}000~deg$^{2}$ of the extragalactic sky in the $g$, $r$, $i$, and $z$ bands across both hemispheres ({{note} that the coverage and number of passes vary between filters; see \url{https://www.legacysurvey.org/dr10/description/} {accessed on 17 June 2024}}). The data have a pixel scale of $0.262''$~pixel$^{-1}$ and a point spread function (PSF) full width at half maximum (FWHM) ranging from $0.6''$ to $1.5''$, with a typical value of $\sim1.1''$~\citep{2020SPIE11447E..94M}. DESI DR10 provides the largest overlap with our samples, including 5617 galaxies from the EGIS catalog imaged in all four bands and 14{,}284 galaxies from the EGIPS catalog. The average $3\sigma$ photometric depths (expressed as surface brightness), measured within a $10\times10$~arcsec$^{2}$ region, reach $29.3 \pm 0.2$, $28.6 \pm 0.2$, $28.2 \pm 0.3$, and $27.7 \pm 0.2$~mag\,arcsec$^{-2}$ in the $g$, $r$, $i$, and $z$ bands, respectively. The distribution of photometric depth is shown in Figure~\ref{fig:desi_sb_egis}. Notably, the $r$-band surface brightness depth of DESI is comparable to that of Stripe~82 ($28.6$~mag\,arcsec$^{-2}$ in the co-added $r$ band), enabling a direct comparison between our results and those of \citet{Stripe82Paper}.

The HSC-SSP survey comprises three layers---Wide (1400~deg$^2$, $r \sim 26$~mag at $5\sigma$), Deep (27~deg$^2$, $r \sim 27$~mag), and UltraDeep (3.5~deg$^2$, $r \sim 28$~mag). The data have a pixel scale of $0.168''$~pixel$^{-1}$ and a PSF FWHM of $\sim$$0.6''$, with a typical range of $\sim$$0.4''$--$0.8''$ depending on observing conditions. From this dataset, we identify 620 galaxies in the $g$, $r$, and $i$ bands that overlap with the EGIS sample. The superior depth and angular resolution of the coadded HSC imaging enhance our sensitivity to faint, extended structures, reaching average photometric depths of $30.3 \pm 0.4$, $29.9 \pm 0.3$, and $29.6 \pm 0.4$~mag\,arcsec$^{-2}$ in the $g$, $r$, and $i$ bands, respectively.

For a subset of particularly interesting galaxies in the EGIS and EGIPS catalogs, we obtained deep optical imaging with the ARC 3.5-m telescope at Apache Point Observatory (APO) using the ARCTIC instrument, which provides a $7.85$~arcmin$^{2}$ field of view and a pixel scale of $0.228''$~pixel$^{-1}$. Targets were selected based on the presence of faint structures that appeared ambiguous or only marginally detectable in shallower imaging. To date, 25 galaxies have been observed, with 14 overlapping the EGIS sample (see Table~\ref{table:table1} for observational details). The data were obtained in the SDSS $g$ and $r$ bands (two galaxies in $g$ only and one in $r$ only) and reach average photometric depths of $29.4 \pm 0.3$ and $29.0 \pm 0.5$~mag\,arcsec$^{-2}$ in the $g$ and $r$ bands, respectively.

Although the APO sample is small, its increased depth significantly improves the detection and characterization of faint tidal features for galaxies not captured by HSC-SSP. These observations highlight the critical role of photometric depth in identifying LSB structures and provide an important validation of our overall results.
\FloatBarrier
\FloatBarrier

\begin{table}[H]
\footnotesize
\caption{Summary of the observational details for galaxy images obtained with APO. The filter name is provided as a {subscript.} 
}

\begin{adjustwidth}{-1.5cm}{-1.5cm}

    \begin{tabularx}{\fulllength}{lCCcCCcc}
\toprule
        \textbf{Name} & \textbf{Date}$_\textbf{\emph{g}}$ & \textbf{Date}$_\textbf{\emph{r}}$ & \textbf{Time} $\boldsymbol{\times}$ $\textbf{\emph{N}}_\textbf{\emph{g}}$ & \textbf{Time} $\boldsymbol{\times}$ $\textbf{\emph{N}}_\textbf{\emph{r}}$ & \textbf{FWHM}$_\textbf{\emph{g}}$ & \textbf{FWHM}$_\textbf{\emph{r}}$\\ \midrule
        EON\_5.686\_14.950 & {12 Oct 2023} & 12 Oct 2023 & $900\times5$ & $900\times4$ & 2.2 & 1.2 \\ 
        EON\_9.910\_14.664 & 12 Oct 2023 & 12 Oct 2023 & $900\times5$ & $900\times4$ & 2.2 & 1.7 \\ 
        EON\_46.480\_-8.019 & 3 Jan 2024 & 12 Oct 2023 & $900\times4$ & $900\times5$ & 2.4 & 2.8 \\ 
        EON\_117.934\_42.253 & N/A & 3 Jan 2024 & N/A & $900\times4$ & N/A & 2.4 \\ 
        EON\_137.491\_43.851 & 28 Feb 2022 & 28 Feb 2022 & $900\times5$ & $900\times5$ & 1.4 & 1.8 \\
        EON\_146.725\_23.023 & N/A & 15 Nov 2023 & N/A & $900\times4$ & N/A & 1.7 \\ 
        EON\_148.062\_47.076 & 25 Mar 2022 & 25 Mar 2022 & $900\times5$ & $900\times5$ & 1.3 & 1.7 \\ 
        EON\_164.403\_44.071 & 25 Mar 2022 & 25 Mar 2022 & $900\times5$ & $900\times5$ & 2.0 & 2.0 \\ 
        EON\_184.615\_12.696 & 5 May 2022 & 5 May 2022 & $900\times5$ & $900\times5$ & 2.0 & 1.9 \\ 
        EON\_191.925\_26.982 & 28 Mar 2022 & 28 Mar 2022 & $900\times5$ & $900\times5$ & 1.3 & 1.4 \\ 
        EON\_201.959\_40.080 & 5 May 2022 & 5 May 2022 & $900\times5$ & $900\times5$ & 1.8 & 1.6 \\ 
        EON\_210.947\_35.742 & 28 Mar 2022 & 28 Mar 2022 & $900\times5$ & $900\times5$ & 1.1 & 1.3 \\
        EON\_235.731\_9.228 & 5 May 2022 & 5 May 2022 & $900\times5$ & $900\times5$ & 1.8 & 2.0 \\ 
        EON\_330.124\_11.380 & 5 May 2022 & 5 May 2022 & $900\times5$ & $900\times5$ & 1.4 & 1.2 \\ 
        \bottomrule
        
    \end{tabularx}
\end{adjustwidth}
\noindent{\footnotesize{{Notes:} 
Some galaxies were observed only in one of the bands. The first column lists the galaxy names as given in the EGIS catalog. The second and third columns indicate the observation dates for each band. The fourth and fifth columns provide the exposure time (in seconds) and the number of exposures. The sixth and seventh columns list the PSF FWHM values for the target galaxy in each band, measured in arcsec.}}
\label{table:table1}
\end{table}


\section{Image Preparation}
\label{sec:imgprep}

To classify LSB features in our samples, we prepared and systematically processed FITS images obtained from DESI, HSC-SSP, and APO datasets prior to visual inspection. The processing workflow was designed to maximize the detectability of faint diffuse structures while preserving the morphological integrity of the targets. For each dataset, image calibration and enhancement procedures were optimized according to the specific instrumental characteristics, depth, spatial resolution, and background properties of the observations. The resulting processed images were then used for consistent visual classification of LSB features across the entire sample. The image preparation methodology adopted for each data source is described in detail below.

\subsection{DESI}
\label{subsec:desi}

Following \citet{2010AJ....140..962M}, we constructed composite images designed to preserve the intrinsic RGB appearance of the central galaxies while simultaneously enhancing the visibility of faint LSB structures in the surrounding background. The image processing pipeline was optimized to improve the detectability of diffuse tidal features, stellar streams, shells, and extended halos.

Initially, the FITS images were geometrically aligned through rotation and subsequently cropped to a uniform field centered on each target galaxy. Source masks were then applied to isolate the primary galaxies and suppress contamination from neighboring objects and foreground stars. To increase the signal-to-noise ratio of faint extended emission, the $g$, $r$, and $i$ (if available) band images were coadded to produce a deep monochromatic background frame.

For the DESI DR9 and DR10 imaging data, an automated sky-background subtraction has already been applied within the survey reduction pipeline. However, visual inspection revealed that, in a subset of cases, the subtraction in DR10 was excessively aggressive and resulted in the partial removal of genuine diffuse emission. In such cases, we reverted to the corresponding DR9 imaging products, which retain a more conservative background treatment in many cases. The coadded background images were subsequently inverted and smoothed in order to enhance low-contrast structures and improve their visibility during visual classification.

Color RGB representations of the central galaxies were generated using the Legacy Sky Viewer visualization {code} 
 ({\url{https://github.com/legacysurvey/imagine/blob/c9cb5d5cbe6023d5d320d2399f98474a51289573/map/views.py\#L5383}} {accessed on 10 May 2024)} and overlaid onto the processed background frames. This approach preserves the natural color morphology of the galaxies while simultaneously emphasizing faint diffuse features in the surrounding regions.

Following visual inspection and quality assessment, images affected by severe artifacts, corrupted regions, or incomplete imaging coverage were excluded from further analysis. The final processed dataset contained 5606 images in the EGIS catalog. An identical image-processing and selection procedure was applied to the EGIPS sample, resulting in a final set of 14{,}237 processed images.

\subsection{HSC-SSP}
\label{subsec:hsc}

The HSC-SSP images were processed using the same general pipeline as adopted for the DESI data to ensure methodological consistency across the datasets. To facilitate automated retrieval and visualization of the images, we incorporated and modified the official HSC-SSP color-postage generation code ({\url{https://hsc-gitlab.mtk.nao.ac.jp/ssp-software/data-access-tools/-/blob/master/pdr3/colorPostage/colorPostage.py}} {accessed on 1 August 2024}) within our processing framework. This implementation was used to obtain calibrated FITS images in the $g$, $r$, and $i$ bands and to generate RGB composite images that were overlaid on the processed LSB background images to facilitate the identification of faint structures.

As with the DESI imaging, the HSC-SSP data required careful treatment of the sky background. During processing of the PDR3 images, we identified a subset of objects affected by overly aggressive sky subtraction, which partially removed diffuse emission and obscured the outer LSB regions of the galaxies. Because a uniform reprocessing of the background subtraction across the full dataset was computationally impractical, we supplemented the affected sample with corresponding images from PDR2, which generally preserved extended LSB structures more effectively, owing to a less aggressive background treatment.

The resulting coadded background images were processed in the same manner as the DESI images, including inversion and smoothing operations to enhance faint diffuse structures. RGB galaxy representations generated from the HSC-SSP imaging were then overlaid onto the processed backgrounds to preserve the intrinsic color of the galaxies while emphasizing surrounding LSB features.

Following visual assessment, images exhibiting severe artifacts, incomplete imaging coverage, substantial residual background contamination, or poor sky subtraction were excluded from the final sample. After this filtering procedure, the final HSC-SSP dataset consisted of 597 processed images derived from a combination of PDR2 and PDR3 observations. 

\subsection{{APO} 
}
\label{sec:apo}

The raw APO FITS images were reduced using the IMAN package ({\url{https://bitbucket.org/mosenkov/iman_new/src/master/}} {accessed on 1 January 2025)} following standard optical image-reduction procedures, including bias subtraction, flat-field correction, cosmic-ray removal, and sky background subtraction. The reduced images were subsequently processed using the same methodology adopted for the DESI and HSC-SSP datasets in order to ensure a homogeneous treatment of all imaging data used in this study. This processing included image alignment, cropping, masking of contaminating sources, coaddition of the available bands to enhance the signal-to-noise ratio of faint diffuse emission, and subsequent inversion and smoothing of the background frames to improve the visibility of LSB structures.

{Since} 
the APO observations were obtained in only two photometric bands, the dataset did not provide sufficient color information to construct reliable intrinsic RGB representations of the galaxies. Therefore, RGB overlays for the APO sample were generated using the corresponding DESI imaging products. This approach enabled consistent visualization of the central galaxy morphology across all datasets while preserving the enhanced LSB background structures detected in the APO observations.

\section{Visual Classification of LSB Tidal Features}
\label{sec:classification}

In this section, we describe the methodology used to identify and classify LSB tidal features in our galaxy samples. We first define the taxonomy of tidal structures based on their morphological characteristics as established in the literature. We then outline the visual classification procedure, including the voting scheme and confidence levels assigned to detected features, and present the resulting incidence of tidal structures in the EGIS and EGIPS samples. In addition, we use APO observations for galaxies with distinctive LSB features to characterize their morphologies and reveal any new LSB features that are not seen in their DESI images. Finally, we assess the impact of galactic cirrus contamination on our classifications.

\subsection{The Taxonomy of LSB Structures}
\label{sec:lsb_taxonomy}

\begin{figure}[H]
    \begin{adjustwidth}{-1.5cm}{-1.5cm}
    \isPreprints{\centering}{} 
    \centering
    \includegraphics[scale = 0.65]{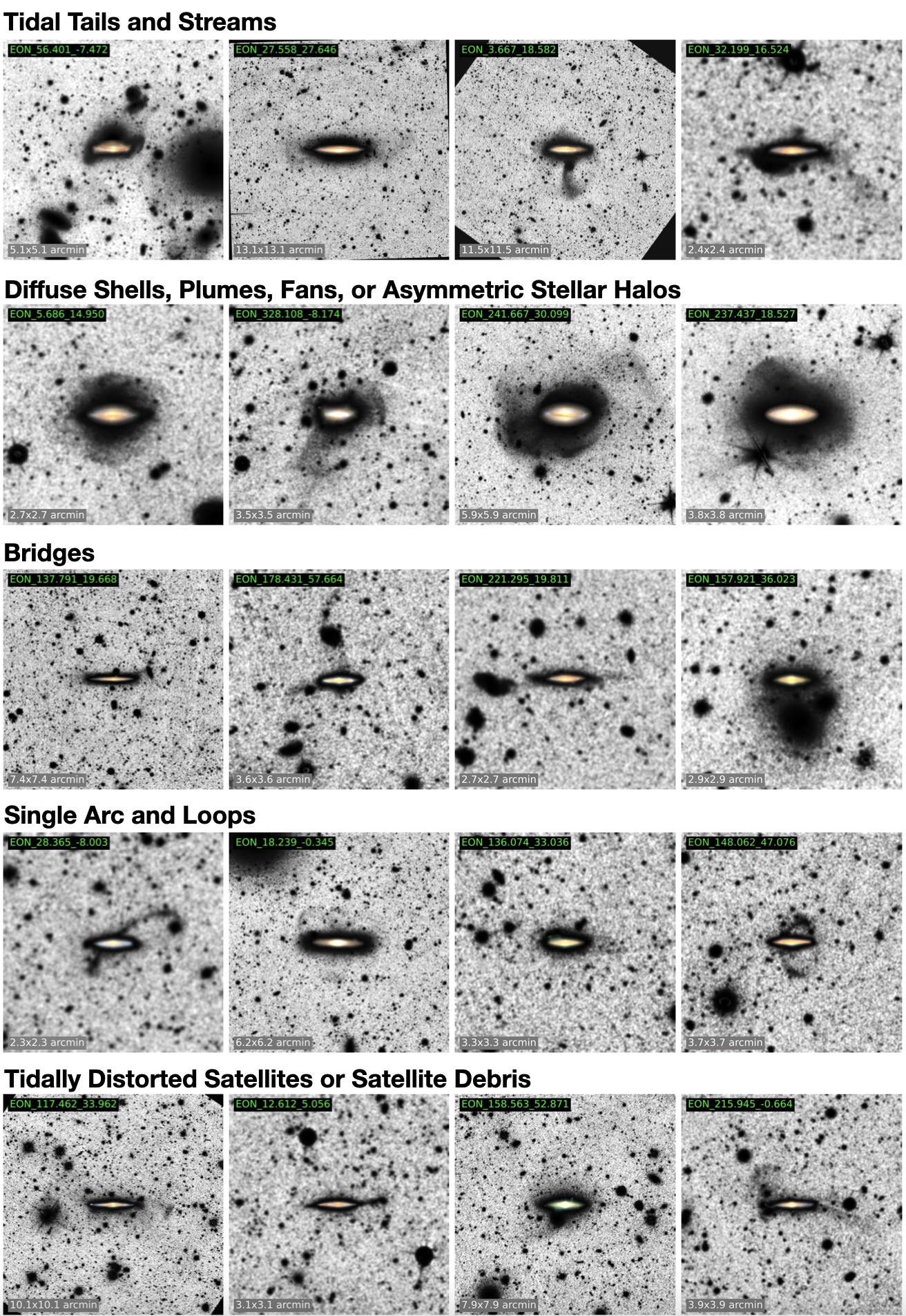}
    \end{adjustwidth}
\caption{Examples of galaxies with different types of LSB tidal features in the EGIS catalog. The target galaxy region enclosed by the $r$-band 25~mag\,arcsec$^{-2}$ isophote shows a DESI RGB composite ($grz$), while the surrounding area displays an enhanced coadded image constructed from the $g$, $r$, and $i$ bands (where available), emphasizing the faint tidal structures. All the images were rotated so that the galaxy's major axis is oriented horizontally.}
    \label{fig:EGIS_examples}
\end{figure}

\begin{figure}[H]
    \begin{adjustwidth}{-1.5cm}{-1.5cm}
    \isPreprints{\centering}{} 
    \centering
    \includegraphics[scale = 0.65]{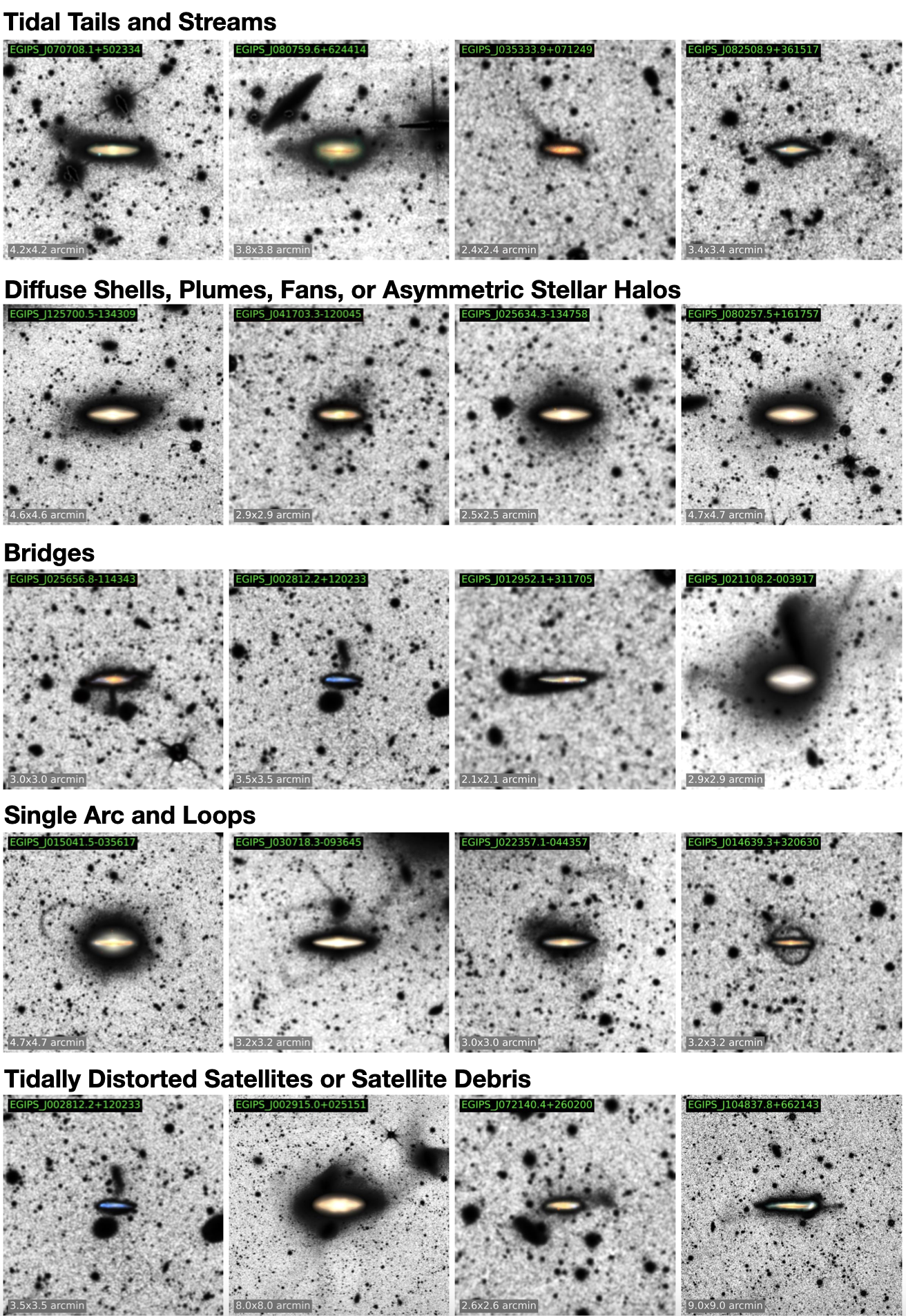}
    \end{adjustwidth}
\caption{Examples of galaxies exhibiting different types of LSB tidal features in the EGIPS catalog. The image presentation is identical to that of Figure~\ref{fig:EGIS_examples}.}
    \label{fig:EGIPS_examples}
\end{figure}

In this work, we focus on LSB features associated with gravitational interactions between galaxies, hereafter collectively referred to as tidal structures. In deep imaging data, tidal features typically appear as faint, diffuse structures surrounding galaxies and often exhibit complex, irregular, or asymmetric morphologies. These diffuse stellar structures are produced by tidal forces during galaxy encounters, mergers, and accretion events, and therefore provide important observational tracers of the hierarchical assembly history of galaxies. Because tidal structures can remain visible for several gigayears after an interaction event~\citep{2004IAUS..217..390M,2013LNP...861..327D,2019A&A...632A.122M,2024MNRAS.529..810R}, they preserve valuable information about the dynamical evolution, merger history, orbital geometry, and mass ratios of interacting systems. Examples of tidal structures identified in the EGIS and EGIPS samples are presented in Figures~\ref{fig:EGIS_examples} and \ref{fig:EGIPS_examples}.

In this study, tidal structures are classified into several morphological categories commonly associated with galaxy interactions and merger activity. The classification is intended to provide a phenomenological description of the observed LSB features rather than a direct inference of their physical origin. The morphology and detectability of tidal structures depend on numerous factors, including the interaction stage, viewing geometry, stellar mass ratio, orbital configuration, gas content, angular resolution, and imaging depth. 
As a result, the classification of tidal structures is not always straightforward. Many features display characteristics of multiple morphological classes, while their apparent morphology can be strongly affected by projection effects and surface-brightness sensitivity. Consequently, the adopted classification scheme is inherently subjective, with some overlap between categories and an expected degree of variation among independent classifiers.

Given the inherent ambiguity and overlap between morphological classes, as well as the dependence of feature appearance on projection effects and image depth, we do not analyze the individual categories separately. Instead, throughout this work, we focus on the overall occurrence of LSB tidal features and consider only their combined statistics. The adopted morphological classes are listed below (see Figures~\ref{fig:EGIS_examples} and \ref{fig:EGIPS_examples} for examples of galaxies from the EGIS and EGIPS samples illustrating each LSB feature type).\

\begin{itemize}[label={}, leftmargin=*,leftmargin=0.0em,labelsep=0.0mm]
  \item\textbf{{Tidal Tails and Streams} 
} 
\end{itemize}

In deep optical images, tidal tails and stellar streams appear as extended filamentary structures emanating from galaxies. Although they can appear similar morphologically, they have distinct origins and physical properties. Tidal tails are typically broad, fan-like extensions of stars and gas originating from galactic disks, often appearing as symmetric pairs in interacting systems~\citep{1972ApJ...178..623T,1992ApJ...393..484B,2019A&A...632A.122M}. They are primarily produced during strong galaxy interactions and major mergers~\citep{1992AJ....103.1089B,2013LNP...861..327D,2015ApJ...807...73O,2020MNRAS.498.2138B}. In optical imaging, tidal tails often exhibit blue, clumpy morphologies due to ongoing star formation within dense condensations, although they also contain older stellar populations stripped from the disk~\citep{2023MNRAS.526.2341R}.

In contrast, stellar streams are narrow, coherent filaments of debris that may wrap partially or fully around a host galaxy. They form through the tidal disruption of low-mass satellite galaxies~\citep{1996ApJ...465..278J,1999AJ....118.1709M,McConnachie_2009,duc14}. Streams are typically more metal-poor and lack significant star formation, reflecting their origin in older stellar populations~\citep{duc14}. They are also generally longer-lived than tidal tails~\citep{2019A&A...632A.122M}.

\begin{itemize}[label={}, leftmargin=*,leftmargin=0.0em,labelsep=0.0mm]
  \item\textbf{{Diffuse Shells, Plumes, Fans, or Asymmetric Stellar Halos}}\end{itemize}

Diffuse shells appear as faint, concentric, sharp-edged arcs or ripples, most commonly observed in early-type galaxies (ETGs)~\citep{2013arXiv1312.1643E,2018MNRAS.480.1715P,2024ApJ...965..158Y}. These structures are generally attributed to galaxy interactions, including both major and minor mergers~\citep{2013arXiv1312.1643E,2018MNRAS.480.1715P}. In the case of minor mergers, shells are thought to arise from the accretion of low-mass satellites on nearly radial orbits~\citep{1990dig..book...72P,2011ApJ...743L..21C,2013arXiv1312.1643E,2018MNRAS.480.1715P}.

Plumes are broad, diffuse stellar structures that extend asymmetrically from the main body of a galaxy, lacking the regular, concentric geometry characteristic of shells, although the two may be physically related. Asymmetric stellar halos, in contrast, are defined by the absence of distinct substructures such as streams, tails, or shells, and instead exhibit large-scale asymmetries in the stellar distribution~\citep{Martin2022,2024MNRAS.530.4422K}.

Fans are broad, diffuse stellar structures extending from the main body of a galaxy and characterized by a wedge-shaped morphology. Although they resemble shells in appearance, they are generally less spatially extended and subtend smaller opening angles. They are generally associated with galaxy interactions and mergers and are often observed in systems exhibiting other forms of tidal debris, suggesting an origin in the redistribution of stars during merger events~\citep{2013ApJ...765...28A,2013arXiv1312.1643E}. Owing to their diffuse nature and often irregular boundaries, fans can be difficult to distinguish from plumes or asymmetric stellar halos, and their classification may depend on imaging depth and viewing angle.

\begin{itemize}[label={}, leftmargin=*,leftmargin=0.0em,labelsep=0.0mm]
  \item\textbf{{Bridges}}\end{itemize}

Bridges are elongated tidal structures composed of stars and gas that extend between interacting galaxies, physically connecting the interacting galaxies and enabling the transfer or redistribution of material between them. These features are generated by tidal forces during close gravitational encounters and are commonly observed in interacting pairs and ongoing mergers. Bridges are frequently more prominent in H{\sc I} observations than in optical imaging, reflecting their substantial gaseous component, and may contain localized regions of active star formation triggered by tidally induced gas compression~\citep{1997ApJ...483..754S,2003AJ....125.1897M}. The formation and morphology of tidal bridges depend sensitively on the orbital geometry, encounter velocity, and mass ratio of the interacting galaxies. Such structures can arise during both close flybys and merger events, particularly in interactions involving galaxies of comparable mass, where strong tidal perturbations may simultaneously produce additional features such as tidal tails and extended diffuse debris~\citep{1972ApJ...178..623T,2010AJ....139.1212S,2017MNRAS.472.2554B}.

\vspace{30pt}
\begin{itemize}[label={}, leftmargin=*,leftmargin=0.0em,labelsep=0.0mm]
  \item\textbf{{Single Arcs and Loops}}
  \end{itemize}

Arcs and loops are coherent LSB stellar structures that can be regarded as specific morphological manifestations of stellar streams, frequently appearing as partially or fully wrapped features surrounding the host galaxy. These structures are produced through the tidal disruption of satellite galaxies during accretion events, where tidal stripping and dynamical friction progressively redistribute stars along the satellite orbit~\citep{2008MNRAS.383...93B,2008ApJ...673..226P}. As with the other types of LSB tidal structures discussed above, the observed morphology depends strongly on the orbital configuration, viewing angle, and evolutionary stage of the interaction. Consequently, the resulting debris may appear as narrow arcs, shell-like features, or extended looping structures surrounding the host galaxy. Because such features can persist over long dynamical timescales, they provide valuable tracers of the recent accretion history and assembly processes of galaxies. Arcs and loops are therefore widely interpreted as observational signatures of past or ongoing minor mergers and satellite accretion events~\citep{2010AJ....140..962M}.

\begin{itemize}[label={}, leftmargin=*,leftmargin=0.0em,labelsep=0.0mm]
  \item\textbf{{Tidally Distorted Satellites or Satellite Debris}}  \end{itemize}

Tidally disrupted satellites are the remnants of accreted satellite galaxies that have undergone partial or extensive tidal disruption by their host galaxy~\citep{1998ApJ...508L..55M,1999AJ....118.1709M,2001ApJ...557..137J,2012Natur.482..192R,2018MNRAS.478.3879S}. As with stellar arcs and loops, they are observational manifestations of stellar streams, representing different stages and geometries of the tidal stripping process. During accretion, tidal forces progressively strip stars, gas, and dark matter from the infalling companion, producing diffuse asymmetric structures, disrupted stellar envelopes, and extended debris fields surrounding the host galaxies. Depending on the stage of disruption and the orbital configuration of the satellite, these features may appear as elongated stellar streams, diffuse overdensities, irregular debris structures, or partially disrupted companions retaining a distorted morphology and, occasionally, a recognizable stellar core. Tidally disrupted satellites represent direct observational evidence of hierarchical galaxy assembly and ongoing minor merger activity, and they provide important constraints on the dynamical evolution of satellite systems and the buildup of stellar halos in massive galaxies.

\subsection{Classification and Incidence of LSB Tidal Features in EGIS and EGIPS}
\label{sec:lsb_classification}

LSB features, classified according to the taxonomy described in Section~\ref{sec:classification}, were identified using a methodology similar to that of \citet{Stripe82Paper}, based primarily on visual inspection of processed galaxy images from DESI and, in the case of the EGIS sample, supplemented by HSC-SSP imaging where available. Visual classification remains one of the most effective approaches for identifying faint, diffuse tidal structures, whose complex and often highly irregular morphologies are difficult to characterize reliably using fully automated methods, including convolutional neural networks (CNNs; see the discussion in \citet{2026arXiv260102579B}). The classification procedure was designed to ensure both consistency and reliability across the full samples while minimizing biases associated with individual classifiers.

The classification team consisted of both experienced researchers (hereafter, experts) and trained student classifiers. Experts were defined as individuals with several years of experience in the identification and analysis of galactic structures, relevant peer-reviewed publications in the field, and prior involvement in training and supervising visual classification efforts. Student classifiers had limited prior experience in identifying LSB structures but underwent dedicated training for this project. The training process included instruction on the morphology and interpretation of tidal features, guided classification exercises using representative examples, and repeated comparison with expert classifications in order to establish consistent identification criteria across the team.

During the visual inspection process, classifiers examined the processed images for evidence of diffuse tidal structures, including tails, streams, loops, bridges, shells, and tidally disrupted satellites.
Particular attention was given to distinguishing genuine astrophysical structures from imaging artifacts, background fluctuations, scattered light, galactic cirrus (see Section~\ref{sec:galCirrus}), and residual processing effects (if still present in the reviewed images). 

\noindent Ambiguous cases and galaxies exhibiting uncertain classifications were reviewed collectively and compared against expert assessments to improve classification consistency and reduce systematic biases. During this validation stage, images with ambiguous morphologies, conflicting votes, or potential contamination from imaging artifacts were re-examined, and misclassifications were corrected where necessary.

The total numbers and corresponding fractions of high-confidence tidal structures are summarized in Table~\ref{tab:class_results}. Because multiple tidal features may be identified within a single galaxy, the total number of classified structures exceeds the total number of galaxies hosting tidal features. In {Appendix} 
 \ref{app} (see Figures~\ref{fig:egis_lsb_sample} and~\ref{fig:egips_lsb_sample}), we present the initial portions of the EGIS and EGIPS catalogs containing the identified LSB structures for the galaxies in our samples, while the complete versions of these catalogs, including all classified objects and associated feature identifications, are provided in the {Supplementary Material} 
 accompanying this paper at \url{https://doi.org/10.5281/zenodo.20573582}.

For the EGIS sample, tidal structures were identified in 324 of 5606 galaxies, corresponding to a tidal feature fraction of 5.8\%. In the EGIPS sample, tidal structures were detected in 690 of 14{,}237 galaxies, yielding a somewhat lower fraction of 4.8\%. This modest difference likely reflects variations in sample selection and intrinsic galaxy properties. These factors, together with the implications of the classification results and their dependence on observational limitations, are discussed in Section~\ref{sec:discussion}.

Restricting the analysis to the complete subsamples (see Section~\ref{sec:sample}), increases the fraction of galaxies with identified LSB features to approximately 6\% (Table~\ref{tab:lsb_statistics_complete}). This increase is expected because larger angular sizes facilitate the detection of faint tidal structures, whereas such features become progressively more difficult to identify in smaller and typically more distant galaxies.

\begin{table}[H]
   \caption{{Statistics of galaxies with identified tidal structures of each type in the EGIS and EGIPS catalogs. The second column lists the total number of galaxies with LSB features and the corresponding percentage relative to the total number of classified galaxies in each catalog.}}
    \renewcommand{\arraystretch}{1.5}
    \begin{tabularx}{\textwidth}{CCCCCCC}
        \toprule
        \textbf{Catalog} & \textbf{Number (Percent)} & \textbf{Tails and Streams} & \textbf{Shells} & \textbf{Bridges} & \textbf{Arcs and Loops} & \textbf{Satellite Debris} \\
        \midrule
        EGIS & 324 (5.8\%) & 134 & 72 & 49 & 84 & 27\\
        EGIPS & 690 (4.8\%) & 260 & 193 & 153 & 159 & 78\\
\bottomrule
    \end{tabularx}
 
    \label{tab:class_results}
\end{table}

\vspace{-12pt}
\begin{table}[H]
\caption{{Total} 
 numbers of galaxies in the complete samples for each catalog and the corresponding numbers of galaxies exhibiting identified LSB features. Percentages relative to the complete samples are given in parentheses.}
\begin{tabularx}{\textwidth}{CCC}
\toprule
\textbf{Survey} & {\textbf{Complete Sample}} & {\textbf{LSB in Complete Sample}} \\
\midrule
EGIS & 4619 & 295 (6.4\%) \\
EGIPS & 5454 & 338 (6.2\%) \\
\bottomrule
\end{tabularx}

\label{tab:lsb_statistics_complete}
\end{table}

\subsection{{APO LSB Features}}

As described in Section~\ref{sec:apo}, the APO imaging provides an independent and, in many cases, deeper view of the LSB structure surrounding a handful of galaxies in our sample. These data are particularly useful for assessing the morphology of faint tidal features whose appearance may be ambiguous in the DESI images alone. In several cases, the APO observations confirm structures already visible in the DESI data, such as stellar streams, shells, loops, and polar-ring candidates. In other objects, the APO images reveal additional diffuse emission, clarify the continuity of previously disconnected features, or expose faint extensions that alter the interpretation of the galaxy's outer morphology.

In the following paragraphs, we discuss the APO images of the individual galaxies listed in Table~\ref{table:table1}, focusing on the LSB structures detected around each galaxy and their possible physical origins. Particular attention is given to features associated with tidal interactions, minor mergers, satellite accretion, and polar or highly inclined structures. Since all galaxy images were rotated such that their major axes are oriented horizontally, we refer to directions within the images as left, right, top, and bottom rather than using the corresponding celestial cardinal directions. {All morphological galaxy classifications were obtained from HyperLeda~\citep{2014A&A...570A..13M}, while redshifts were retrieved from the NASA/IPAC Extragalactic Database (NED)~\citep{1991ASSL..171...89H}.}

\begin{itemize}[label={}, leftmargin=*,leftmargin=0.0em,labelsep=0.0mm]
  \item\textbf{{EON\_5.686\_14.950}}
  \end{itemize}

EON\_5.686\_14.950, also known as 2MASXJ00224457{+}
1456586, is included in the sample of 54 disk galaxies with prominent stellar halos studied by \citet{2018JKAS...51...73A}.  The galaxy has a redshift of $z=0.025$ and is classified as an Sa galaxy. 
\citet{2018JKAS...51...73A} also noted that 2MASXJ00224457+1456586 is one of the galaxies in their sample whose outer disk exhibits relatively blue colors, indicating the presence of spiral arms. Their two-component decomposition yielded a de Vaucouleurs bulge with a S\'ersic index of $n_{b}=4.0$ and a stellar halo with a S\'ersic index of $n_{h}=1.7$. In the APO image shown in Figure~\ref{fig:apo}, EON\_5.686\_14.950 exhibits a shell-like tidal structure whose morphology is more prominent than in the DESI image. The APO image also reveals a fan-like feature embedded in the bottom part of the shell. The presence of these features may contribute to the relatively high S\'ersic index measured for the stellar halo and is consistent with a past satellite accretion event. This interpretation is in agreement with the broader conclusion of \citet{2018JKAS...51...73A} that prominent stellar halos in disk galaxies are linked to galaxy assembly processes, including the accretion and disruption of satellite galaxies.
\begin{itemize}[label={}, leftmargin=*,leftmargin=0.0em,labelsep=0.0mm]
  \item\textbf{{EON\_9.910\_14.664}}
  \end{itemize}

EON\_9.910\_14.664, also known as 2MASXJ00393835+1439509, is a nearby disk galaxy at a redshift of $z = 0.017$ with a morphological type of Sa~\citep{2010ApJS..186..427N}. In Figure~\ref{fig:apo}, the color image of the galaxy reveals a possible polar bulge. Galaxies hosting polar bulges exhibit a major axis that is significantly inclined relative to the major axis of the disk. Their morphology closely resembles that of polar-ring galaxies; however, polar bulges appear to be less luminous analogs of the host galaxies in polar-ring systems~\citep{2015AstL...41..748R}. Although this type of feature was not included in our primary classification scheme, it illustrates the role of deep imaging in refining morphological interpretations. In particular, the APO image reveals a more lopsided and warped disk than is apparent in the DESI image, as well as a very faint stellar stream extending from a small galaxy located to the left and in close proximity to EON\_9.910\_14.664. These features suggest that the LSB stellar component is more effectively recovered in the deeper APO data. Polar structures, including polar bulges and polar rings, may form through several channels, such as the accretion of material tidally stripped from a companion galaxy~\citep{1983AJ.....88..909S,1997A&A...325..933R}, the disruption of a gas-rich satellite on a nearly perpendicular orbit~\citep{1991wdir.conf..112R,1992ApJ...389L..55K}, major mergers between galaxies of comparable mass~\citep{1997ApJ...490L..37B,1998ApJ...499..635B,2003A&A...401..817B}, and the misaligned accretion of gas from cosmic filaments onto the host galaxy~\citep{2006ApJ...636L..25M,2008ApJ...689..678B}. Deep photometry is therefore important for investigating the formation mechanisms of galaxies hosting polar structures~\citep{2022RAA....22k5003M}.

\begin{itemize}[label={}, leftmargin=*,leftmargin=0.0em,labelsep=0.0mm]
  \item\textbf{{EON\_46.480\_-8.019}}
  \end{itemize}

The edge-on galaxy EON\_46.480\_-8.019, also identified as 2MFGC~02521, is classified as an Sc galaxy and has a redshift of $z = 0.028$. Both the DESI and deeper APO images in Figure~\ref{fig:apo} reveal LSB asymmetries on the left side of the disk, indicating a lopsided morphology. Such lopsidedness is a common feature of disk galaxies and is generally interpreted as evidence of a non-axisymmetric mass distribution. Several mechanisms have been proposed to explain its origin, including tidal interactions with nearby companions, minor mergers, asymmetric gas accretion from the surrounding environment, and perturbations of the dark matter halo potential~\citep{2005A&A...438..507B,2009PhR...471...75J}. The detection of these asymmetric LSB features therefore provides valuable clues to the recent dynamical evolution and interaction history of EON\_46.480\_-8.019.

\begin{itemize}[label={}, leftmargin=*,leftmargin=0.0em,labelsep=0.0mm]
  \item\textbf{{EON\_117.934\_42.253}}
  \end{itemize}

EON\_117.934\_42.253, also known as 2MASXJ07514417+4215129, has a morphological type of Sa and a redshift of $z = 0.041$. In Figure~\ref{fig:apo}, its central region exhibits a vertically extended, classical bulge-like component. The deeper APO image also reveals a faint spiral arm that is not readily apparent in the shallower DESI imaging, indicating that the galaxy retains a more developed disk structure than might otherwise be inferred from its edge-on orientation. Interestingly, the deeper APO image further reveals a possible ring-like structure surrounding another edge-on galaxy located to the right of EON\_117.934\_42.253.

\begin{itemize}[label={}, leftmargin=*,leftmargin=0.0em,labelsep=0.0mm]
  \item\textbf{{EON\_137.491\_43.851}}
\end{itemize}

EON\_137.491\_43.851, also known as WISEAJ090957.90+435104.1, is an Sa galaxy with a redshift of $z = 0.025$. The most striking feature of this galaxy in Figure~\ref{fig:apo} is a prominent stellar stream that extends well beyond the main body of the galaxy and wraps around the host. The APO image reveals that the stream extends farther above the galaxy than is apparent in the DESI image, where it bends to the left and appears to connect to a brighter segment of the stream, suggesting that the tidal feature may form a more continuous structure than previously recognized. The large radial extent and coherent, curved morphology of the stream are consistent with tidal debris produced during the accretion and disruption of a lower-mass satellite galaxy. The remnant core of the disrupted satellite may still be visible just above the galaxy center, near the location where the stellar stream appears to originate.

\begin{itemize}[label={}, leftmargin=*,leftmargin=0.0em,labelsep=0.0mm]
  \item\textbf{{EON\_146.725\_23.023}}
  \end{itemize}

EON\_146.725\_23.023, also identified as UGC~05235, has a redshift of $z = 0.024$ and a morphological classification of Sc. A tidal stream is visible in Figure~\ref{fig:apo}, extending from the upper-left side of the galaxy above the disk. The presence of this structure suggests that the galaxy has likely experienced a recent minor merger event. In the color image, the stream exhibits a faint blue hue, indicating that the disrupted dwarf galaxy may host young stellar populations and thus ongoing or recent star formation. In addition, a blue stellar debris feature is visible below the disk of UGC~05235 on its right-hand side, possibly associated with the observed stellar stream. The right side of the main galaxy's disk also appears warped, further supporting a scenario in which UGC~05235 has undergone a gravitational interaction with the dwarf galaxy.

\begin{itemize}[label={}, leftmargin=*,leftmargin=0.0em,labelsep=0.0mm]
  \item\textbf{{EON\_148.062\_47.076}}
   \end{itemize}

EON\_148.062\_47.076, also known as 2MFGC~07640, has a redshift of $z = 0.037$ and is classified as an Sb galaxy. The most prominent feature of this galaxy in Figure~\ref{fig:apo} is a ring- or loop-like stellar structure oriented nearly perpendicular to the plane of the host galaxy. In the DESI image, this feature appears incomplete; however, the deeper APO image reveals a very faint continuation of the structure on the lower-right side of the galaxy, suggesting that the loop may form a more coherent polar ring. Such polar tidal features are commonly interpreted as tidal debris generated during the accretion and disruption of a lower-mass companion during a minor merger event and represent one of the proposed formation channels for faint polar rings in galaxies.

\begin{itemize}[label={}, leftmargin=*,leftmargin=0.0em,labelsep=0.0mm]
  \item\textbf{{EON\_164.403\_44.071}} \end{itemize}

The Sc galaxy EON\_164.403\_44.071, also identified as MCG~+07-23-005, has a redshift of $z = 0.0343$ and exhibits multiple LSB tidal features in Figure~\ref{fig:apo}. In the DESI image, an arc-like stellar stream is visible along the upper-right edge of the galaxy, suggesting the presence of tidally displaced material in the outer halo. The deeper APO image further reveals fainter and narrower stellar streams extending toward the lower-left side of the galaxy. These diffuse structures may form a tidal bridge between EON\_164.403\_44.071 and the nearby galaxy WISEAJ105739.66+440532.9. The similar redshift of WISEAJ105739.66+440532.9, $z = 0.0341$, supports the possibility that the two galaxies are physically associated rather than being a chance projection on the sky. If this interpretation is correct, the observed streams and possible tidal bridge are consistent with material stripped during an ongoing or recent tidal interaction between the two objects.

\begin{itemize}[label={}, leftmargin=*,leftmargin=0.0em,labelsep=0.0mm]
  \item\textbf{{EON\_184.615\_12.696}}
  \end{itemize}

EON\_184.615\_12.696, also known as FGC~1405, has a morphological classification of Sc and a redshift of $z = 0.025$. A prominent stream-like feature is visible on the upper-right side of the galaxy in Figure~\ref{fig:apo}, while additional diffuse material is present on the left side of the disk. Together, these structures may trace an extended arc or even a ring- or loop-like feature surrounding the galaxy; however, no continuous connection between the two components is clearly detected. Any possible bridge between them remains extremely faint and diffuse, and the APO image does not unambiguously confirm that the upper-right stream and the left-side material belong to a single coherent structure. The deeper APO data do, however, reveal additional LSB material extending toward the lower-left region of the galaxy. The absence of a clear connection raises the possibility that the upper-right structure represents a separate tidal feature or remnant satellite debris rather than part of a continuous arc. In either scenario, the disturbed outer morphology of EON\_184.615\_12.696 is consistent with the aftermath of a minor tidal interaction.

\begin{itemize}[label={}, leftmargin=*,leftmargin=0.0em,labelsep=0.0mm]
  \item\textbf{{EON\_191.925\_26.982}}
    \end{itemize}

The morphology of EON\_191.925\_26.982, alternatively identified as UGC~07959, is consistent with that of a polar-ring galaxy. The host is classified as an Sb galaxy and has a redshift of $z = 0.024$. A prominent ring-like structure is visible in Figure~\ref{fig:apo}, oriented nearly perpendicular to the main plane of the galaxy and wrapping around the host in a configuration characteristic of polar or highly inclined stellar material. The feature also appears slightly tilted with respect to the principal axis of the host, giving the ring a relatively thick projected appearance. The deeper APO image reveals that the ring extends farther above and below the galaxy than is apparent in the DESI image, further supporting its interpretation as an extended polar structure. The oval-shaped feature visible in the lower-right corner of the APO image is an imaging artifact associated with the presence of a nearby bright star and is unrelated to the galaxy.
\begin{itemize}[label={}, leftmargin=*,leftmargin=0.0em,labelsep=0.0mm]
  \item\textbf{{EON\_201.959\_40.080}}  \end{itemize}

EON\_201.959\_40.080, also known as WISEAJ132750.22+400447.7, is an Sb galaxy with a redshift of $z = 0.046$. In the enhanced DESI and color images in Figure~\ref{fig:apo}, a faint fan-like feature is visible on the right side of the galaxy. The deeper APO image reveals a substantially more extended structure on the opposite side; namely, a broad shell-like feature to the left of the galaxy that is not apparent in the shallower imaging. The observed shell structure may have formed as a result of either a minor or major merger, depending on the mass ratio and orbital properties of the accreted companion. In addition, both the DESI and APO images reveal a boxy outer morphology of the galaxy disk, suggesting that the stellar distribution has been dynamically perturbed~\citep{Stripe82Paper}.
\begin{itemize}[label={}, leftmargin=*,leftmargin=0.0em,labelsep=0.0mm]
  \item\textbf{{EON\_210.947\_35.742}} \end{itemize}

EON\_210.947\_35.742, identified as UGC~08984, is an Sa galaxy with a redshift of $z = 0.013$. In Figure~\ref{fig:apo}, the galaxy exhibits a prominent tidal structure that we classify as a candidate polar ring. In the DESI image, the upper component appears to curve sharply and loop back toward the host galaxy, supporting the interpretation that the apparent streams may be connected portions of a larger ring-like feature. The APO image emphasizes the LSB debris more strongly, although the increased prominence of the diffuse emission makes the projected continuity of the structure more difficult to trace. Nevertheless, the overall morphology is consistent with externally acquired material, most plausibly debris from a satellite accreted onto an inclined or nearly polar orbit. Thus, EON\_210.947\_35.742 may represent a galaxy in which tidal debris is organized into a developing polar ring. In addition, the right side of the galaxy exhibits a small arc-like feature in the APO image that is only marginally visible in the DESI image.

\begin{itemize}[label={}, leftmargin=*,leftmargin=0.0em,labelsep=0.0mm]
  \item\textbf{{EON\_235.731\_9.228}} \end{itemize}

EON\_235.731\_9.228, also known as CGCG~078-051, is an Sd galaxy with a redshift of $z = 0.035$. In Figure~\ref{fig:apo}, the galaxy exhibits two prominent stellar features extending from the central region on opposite sides of the galaxy, above and below the main disk in a direction nearly perpendicular to its plane. The upper feature displays an umbrella-like morphology. We classify this object as a candidate polar tidal structure with a fan-like component. The umbrella-like morphology is suggestive of tidal debris produced by the disruption of a satellite galaxy. Similar fan-like and umbrella-shaped structures are frequently associated with accretion events occurring on highly eccentric or nearly radial orbits, which can generate broad, diffuse stellar overdensities extending far into the halo of the host galaxy~\citep{2008ApJ...689..936J,2010AJ....140..962M,2015MNRAS.450..575A,2015MNRAS.454.2472H}.

\begin{itemize}[label={}, leftmargin=*,leftmargin=0.0em,labelsep=0.0mm]
  \item\textbf{{EON\_330.124\_11.380}}\end{itemize}

The galaxy EON\_330.124\_11.380, also known as 2MASXJ22002984+1122489, has a redshift of $z = 0.031$ and is classified as an Sc galaxy. In Figure~\ref{fig:apo}, a prominent loop-like tidal feature is visible around EON\_330.124\_11.380. The loop is already clearly detected in the DESI image, indicating that the tidal debris has a relatively high surface brightness compared to many of the fainter structures in the sample. In the APO image, the outer morphology appears more complex, exhibiting a cartwheel-like configuration with spiral-like structures extending from the edge-on galaxy disk. This morphology is suggestive of a collisional ring system produced by a strong gravitational encounter, in which an intruder galaxy passes through or near the disk of a gas-rich galaxy, generating an outwardly propagating density wave that forms ring-like and spiral structures~\citep{1976ApJ...209..382L,1996FCPh...16..111A}. In this scenario, the edge-on disk galaxy may represent either the perturbed disk galaxy itself or the intruding companion projected against the ring-like debris. The observed morphology therefore points to a dynamically disturbed object resulting from a recent interaction or collision, although deeper imaging and kinematic observations would be required to determine the precise geometry of the encounter.

Taken together, the APO observations demonstrate that many of the LSB features identified in the DESI images represent components of more extended and morphologically complex structures than can be inferred from shallower data alone. In several galaxies, the APO imaging reveals previously undetected shells, plumes, streams, bridges, rings, or diffuse stellar envelopes, while in others it clarifies the continuity and full extent of structures already visible in the DESI images. These results indicate that the outer regions of edge-on galaxies frequently contain faint stellar debris that traces past dynamical evolution and that the morphology of such debris is often only partially recovered in shallower surveys (see also Gilhuly et al. \citep{2020ApJ...897..108G}).

A notable outcome of this analysis is the prevalence of features associated with tidal interactions and accretion events. Streams, shells, loops, tails, and asymmetric stellar envelopes are present throughout both the APO sample and the much bigger EGIS and EGIPS samples, suggesting that the assembly histories of these galaxies have been significantly influenced by the accretion and disruption of lower-mass companions. Several galaxies among the selected APO galaxies with LSB features also exhibit polar tidal or well-shaped structures that are themselves commonly linked to the external acquisition of material. Although the APO sample is small, the recurring presence of these features across galaxies spanning a range of morphological types and redshifts suggests that signatures of hierarchical growth are widespread among nearby disk galaxies.

More broadly, these observations highlight the importance of deep imaging for studies of galaxy evolution. Features that appear isolated, fragmented, or ambiguous in shallower imaging are often revealed by the APO data to be components of larger, coherent structures whose morphology provides stronger constraints on their origin. The APO observations therefore reinforce the conclusion that deep LSB imaging is essential for obtaining a more complete census of tidal debris and for reconstructing the recent interaction and accretion histories of nearby galaxies. In this context, forthcoming wide-field surveys such as the LSST are expected to transform studies of LSB structures by providing unprecedented depth, area coverage, and sensitivity to faint tidal features, enabling statistical investigations of galaxy assembly and accretion processes across large samples of nearby galaxies.

\subsection{The Influence of Galactic Cirrus}
\label{sec:galCirrus}

To account for potential contamination of the galaxy images from optical galactic cirrus, we performed a visual inspection of all DESI fields in the EGIS catalog, specifically searching for diffuse foreground emission. Optical galactic cirrus consists of faint, diffuse, filamentary interstellar dust structures that scatter ambient starlight and can mimic or obscure genuine LSB extragalactic features, making it important to identify and account for in such analyses. To this end, we downloaded larger (10$'$ $\times$ 10$'$) field-of-view RGB images for each galaxy to assess the presence of galactic cirrus (see Figure~\ref{fig:cirrusEx}). All candidate LSB fields were additionally inspected using the Legacy Survey Sky Viewer ({\url{https://www.legacysurvey.org/viewer}} {accessed on 1 October 2025)} to evaluate the impact of cirrus contamination on the identified features.

\begin{figure}[H]
\begin{adjustwidth}{-1.5cm}{-1.5cm}
\isPreprints{\centering}{} 
\centering
\includegraphics[width=1.1\textwidth]{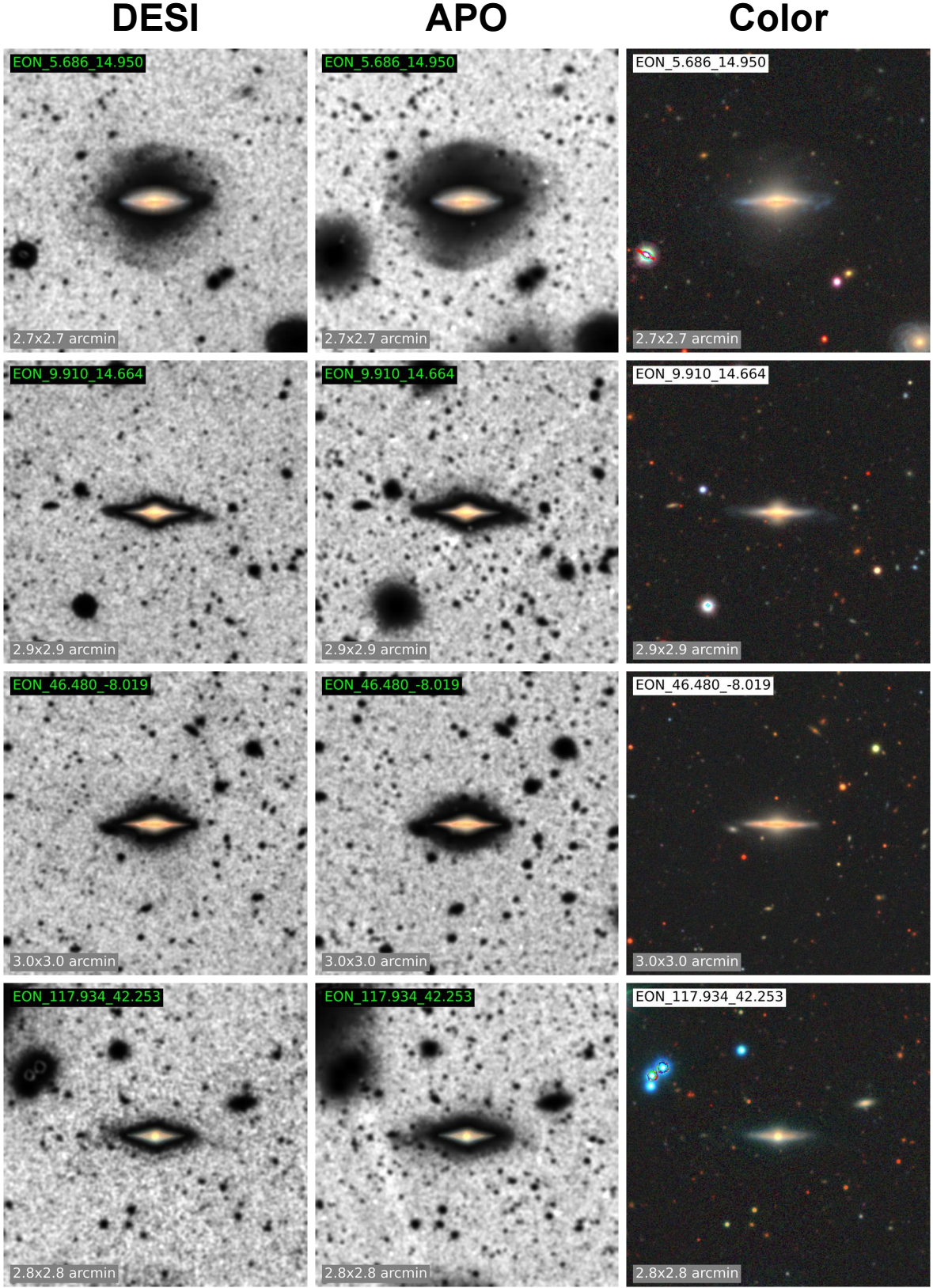}\end{adjustwidth}
\caption{\textit{Cont}.}
\label{fig:apo}

\end{figure}

\begin{figure}[H]\ContinuedFloat

\begin{adjustwidth}{-1.5cm}{-1.5cm}
\isPreprints{\centering}{} 
\centering
\includegraphics[width=1.1\textwidth]{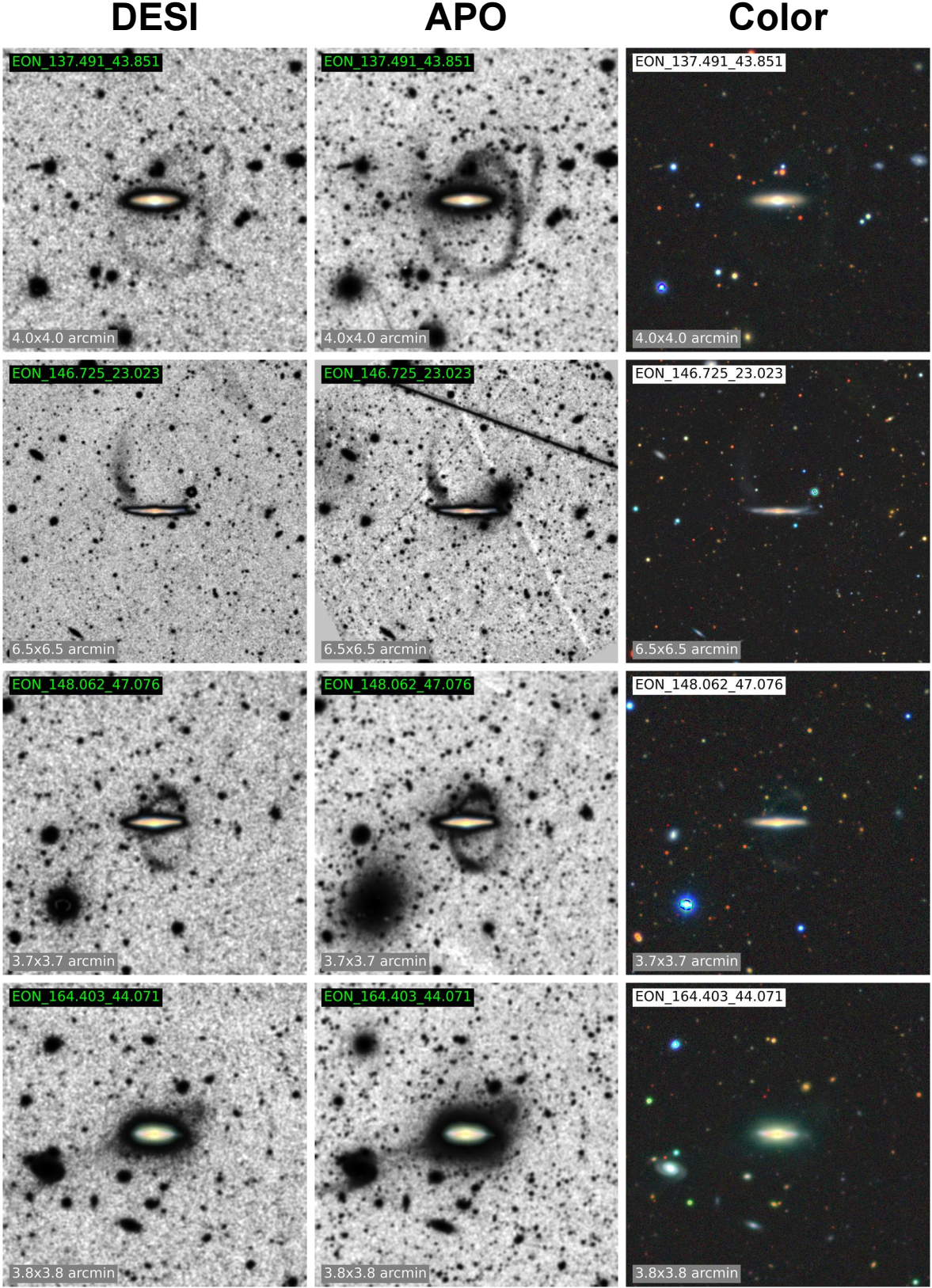}\end{adjustwidth}
\caption{\textit{Cont}.}
\label{fig:apo}

\end{figure}

\begin{figure}[H]\ContinuedFloat
\begin{adjustwidth}{-1.5cm}{-1.5cm}
\isPreprints{\centering}{} 
\centering
\includegraphics[width=1.1\textwidth]{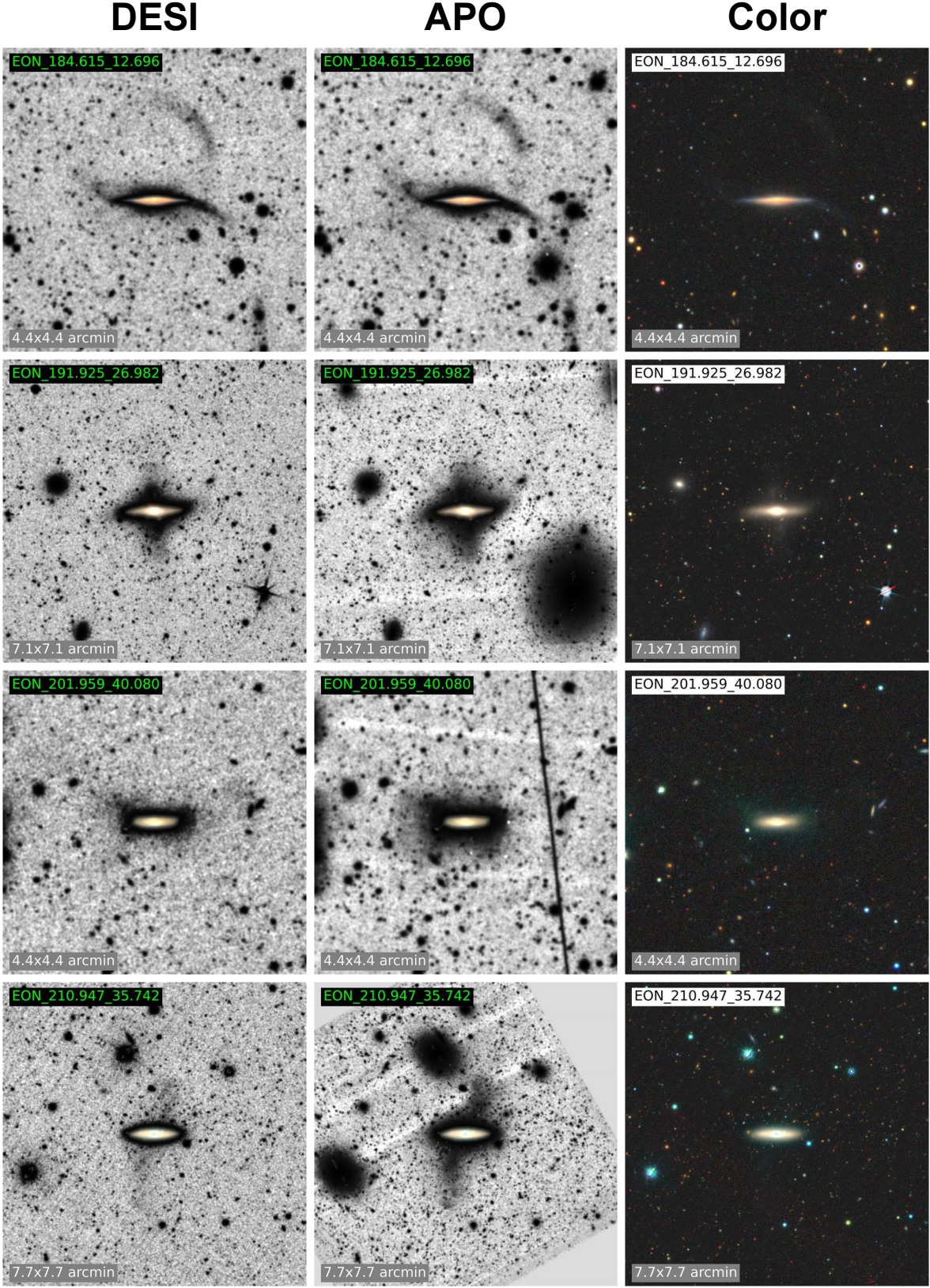}\end{adjustwidth}
\caption{\textit{Cont}.}
\label{fig:apo}

\end{figure}

\begin{figure}[H]\ContinuedFloat
\begin{adjustwidth}{-1.5cm}{-1.5cm}
\isPreprints{\centering}{} 
\centering
\includegraphics[width=1.1\textwidth]{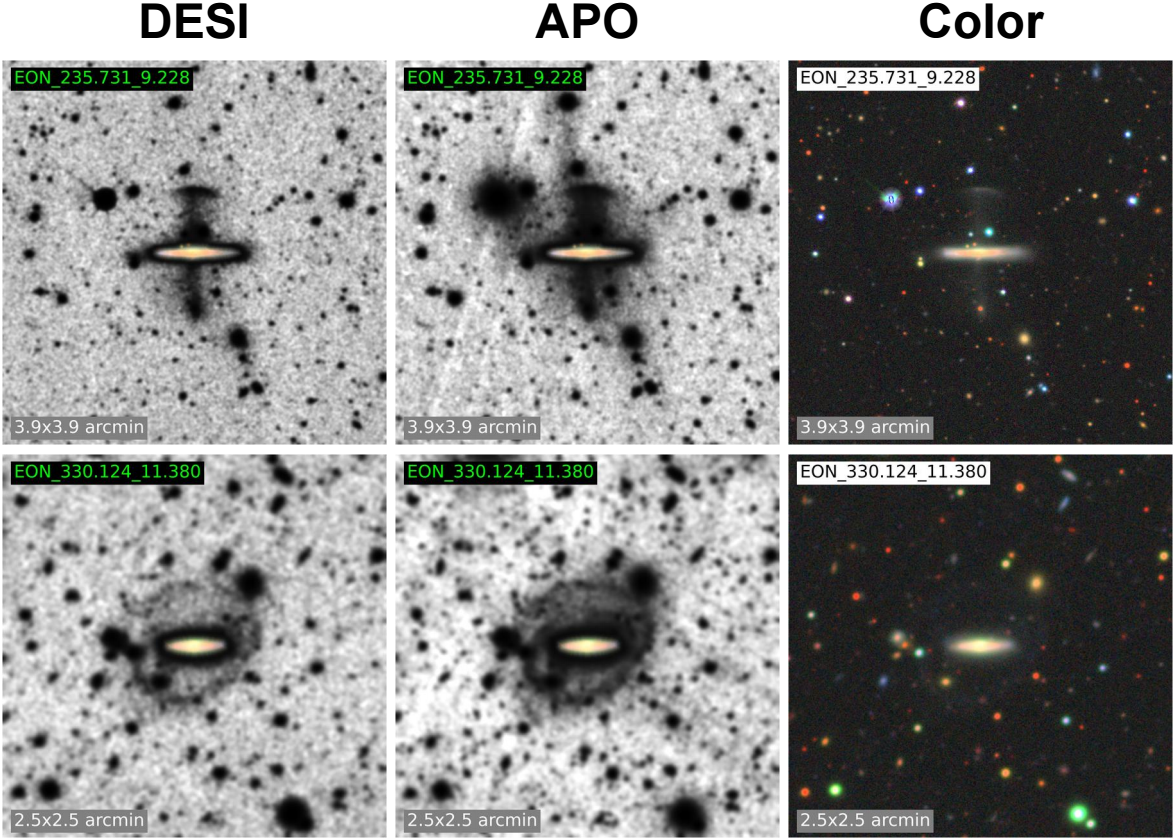}\end{adjustwidth}
\caption{{APO} 
 images presented with DESI and RGB DESI images (created using $g$, $r$, and $z$ bands).}

\end{figure}
\vspace{-6pt}
\begin{figure}[H]
    \begin{adjustwidth}{-1.5cm}{-1.5cm}
    \isPreprints{\centering}{} 
    \includegraphics[scale = 0.674]{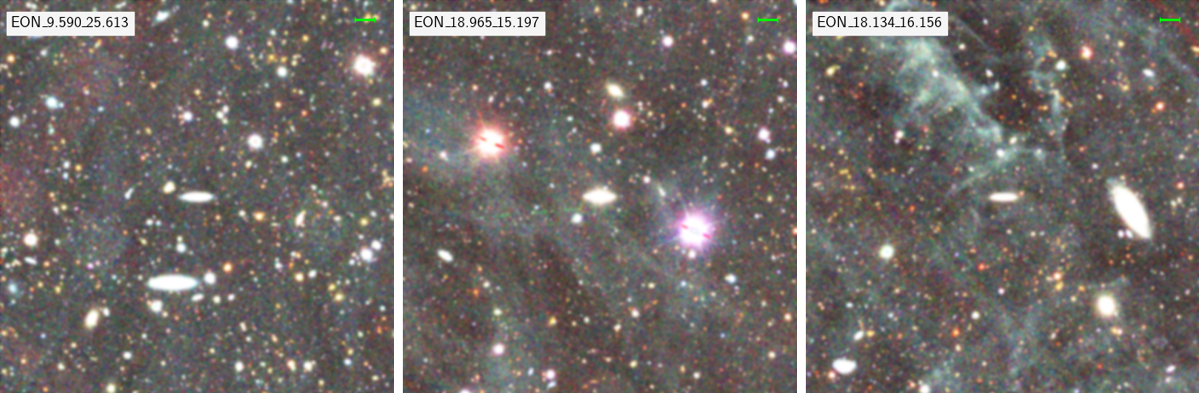}
    \end{adjustwidth}
    \caption{Examples of Galactic cirrus with relative brightness levels classified as dim (left), moderate (middle), and bright (right). Each image is centered on a galaxy from the EGIS catalog. The green bars indicate angular scales of $30''$.}
    \label{fig:cirrusEx}
\end{figure}

Using this procedure, we identified 547 fields (9.5\%) in EGIS affected by cirrus, defined as emission overlapping the galaxy or located within an angular distance comparable to the galaxy radius. These cirrus features were qualitatively classified by relative brightness into three categories: 339 dim, 149 moderate, and 60 bright. Each field was re-examined to ensure classification consistency.

Among the 547 EGIS galaxies affected by galactic cirrus, 11 exhibit tidal features. In these cases, visual inspection indicates that the cirrus contamination is minor and does not compromise the identification of genuine tidal structures. We also note that, during the classification process, galaxies exhibiting clear spatial overlap with cirrus clouds were excluded from further consideration. Such clouds can generally be distinguished from tidal features based on their characteristic morphology and optical colors (see, e.g., Marchuk et al. \cite{2021MNRAS.508.5825M}, Smirnov et al. \cite{2023MNRAS.519.4735S}, Román et al. \cite{2020A&A...644A..42R}). We therefore conclude that galactic cirrus does not significantly affect our classifications.

A similar visual screening was performed for the EGIPS catalog, using only the Legacy Survey Sky Viewer to minimize the risk of misidentifying galactic cirrus as LSB features. A comparable fraction of the fields was found to be significantly affected by galactic cirrus as in the EGIS sample; however, for EGIPS, the cirrus was neither systematically classified nor analyzed in detail.

To further characterize the cirrus-contaminated fields in the EGIS sample, we estimated the corresponding infrared (IR) emission near each galaxy using the IRAS (Infrared Astronomical Satellite, Neugebauer et al.~\cite{1984ApJ...278L...1N}) maps ({\url{https://lambda.gsfc.nasa.gov/product/iras/iras_iris_get.html}} {accessed on 1 September 2022}) at 100~$\upmu$m~\citep{2005ApJS..157..302M}. For each object, we used its sky coordinates to locate its position on the IR map and computed the average intensity from the eight surrounding pixels.

\begin{figure}[H]
\isPreprints{\centering}{} 
 \hspace{-12pt}   \includegraphics[width=12cm]{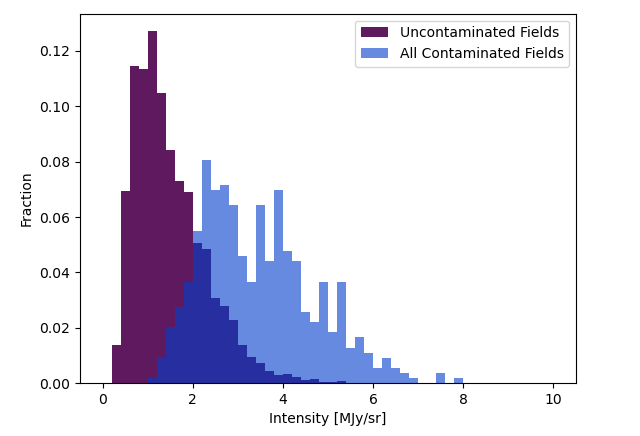}
    \caption{A comparison between IR intensity for galaxies with and without optical galactic cirrus contamination.}
    \label{fig:cirrusHist}
\end{figure}

{Figure}~\ref{fig:cirrusHist} compares the IR intensity distributions for fields with and without optical cirrus contamination. As expected, contaminated fields exhibit systematically higher IR intensities on average, reflecting enhanced cold dust emission. However, the two distributions show substantial overlap. In principle, a clearer separation might be expected, as cirrus contamination should correspond to elevated IR emission from dust grains heated by the interstellar radiation field. The observed overlap between the two distributions can be attributed to several factors. Optical and infrared cirrus do not always correlate directly, as they trace different physical processes and intrinsic dust properties, such as asymmetry and albedo (see e.g., Zhao et al. \cite{2024AJ....168...88Z} and references therein). Optical cirrus is detected via scattered starlight and depends on the illumination geometry and grain characteristics, whereas infrared cirrus arises from thermal emission and is sensitive to dust temperature and emissivity. Consequently, some regions may appear prominent in one wavelength regime but faint or undetectable in the other~\citep{2025ApJ...979..175L}. In addition, differences in spatial resolution, depth, and sensitivity between the optical and infrared datasets can further weaken this correspondence~\citep{2013ApJ...767...80I,2013ApJ...778L..40S,2020A&A...644A..42R,2021MNRAS.508.5825M,2023MNRAS.519.4735S}. These effects likely lead to residual IR emission in fields classified as uncontaminated in the optical, producing the observed overlap and limiting the effectiveness of IR intensity alone as a diagnostic of cirrus contamination.
\vspace{-3pt}

\section{Discussion of Results}
\label{sec:discussion}

In this section, we first discuss the various observational and methodological biases that may influence the results and statistical properties of our classification. We then place our findings in the broader context of previous work by comparing them with results from both observational studies and numerical simulations. Unless stated otherwise, the discussion is based primarily on the EGIS sample, while the EGIPS catalog is used as a supplementary dataset, except in the analysis of redshift and stellar mass distributions, where both samples are considered. Particular attention is given to similarities and discrepancies between our results and those reported in the literature. Potential sources of these differences, including sample selection effects, observational limitations, and differences in classification methodologies, are discussed below.

\subsection{Projection Bias}

Projection effects can influence the visibility of tidal features, as the apparent morphology of a galaxy depends on its viewing angle, which may obscure or distort faint structures. Using the Illustris simulation, \citet{2018MNRAS.480.1715P} showed that only 40\% of galaxies hosting shells are identifiable in all three orthogonal projections, with most shells visible in only two. Similarly, \citet{Martin2022} demonstrated that projection-related uncertainties can be a dominant source of error in deep imaging studies of tidal features. Although such effects cannot be eliminated in observational data, their impact can be mitigated through careful sample selection. In this work, we restricted our analysis to edge-on galaxies, thereby adopting a relatively uniform viewing geometry and reducing variations in the detectability of LSB structures. While projection effects may still affect the visibility and classification of individual features, we expect their influence on the overall tidal fractions derived for our sample to be relatively modest.

\subsection{Classification and Detection Biases}

As described in Section~\ref{sec:classification}, tidal features in our sample were identified through visual classification, a widely used approach in similar studies. While effective, this method is inherently subjective and may introduce biases related to classifier training and individual interpretation, potentially leading to inconsistencies in feature identification~\citep{2010ApJ...709.1067B,2020MNRAS.492.2075B}. Indeed, \citet{Martin2022} demonstrated that even experienced classifiers can disagree when identifying tidal features in mock images, and that classifications may change upon re-evaluation, highlighting the intrinsic uncertainty of this process.

To mitigate these effects, all galaxies in the EGIS sample were examined multiple times by several trained classifiers following consistent guidelines, and the results were independently reviewed by expert classifiers to improve reliability. However, the use of a single research group introduces the possibility of shared biases, which are difficult to quantify directly. To estimate this uncertainty, we adopt the $\sim$$3\%$ disagreement rate reported by \citet{2010ApJ...709.1067B} from a blind classification study, treating it as a lower limit on the uncertainty in our classifications. Although subjectivity cannot be fully eliminated, these steps help to reduce its impact and enhance the robustness of our results.

\begin{figure}[H]
\begin{adjustwidth}{-1.5cm}{-1.5cm}
\centering
\subfloat[\centering]{\includegraphics[width=7cm]{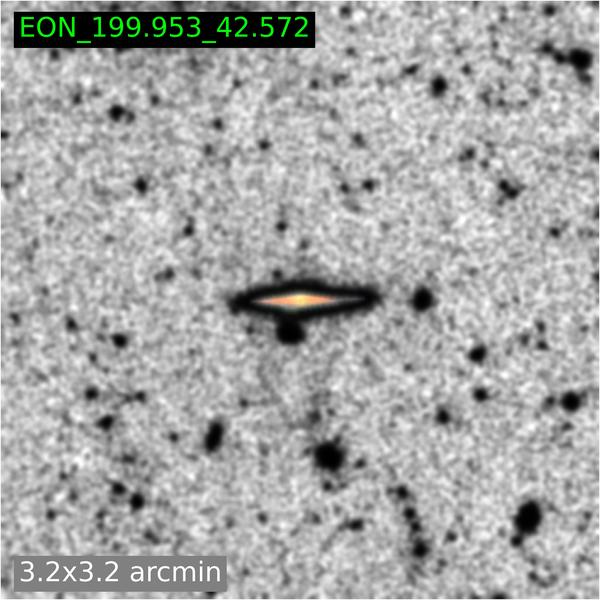}}
\subfloat[\centering]{\includegraphics[width=7cm]{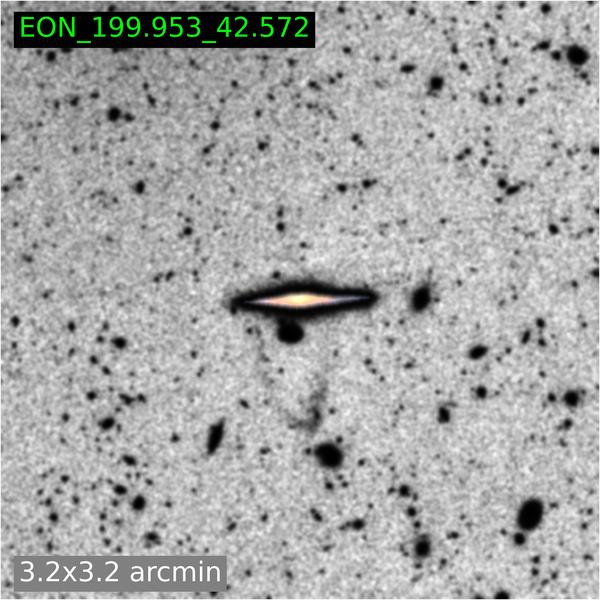}}\\
\end{adjustwidth}
\caption{Comparison of image depth for the same galaxy in DESI (\textbf{a}) and HSC-SSP (\textbf{b}). A faint tidal stream is visible in the HSC-SSP image as an arc extending below the galactic plane, while it remains undetected in the shallower DESI data.}
  \label{fig:desi_hsc}
\end{figure}

The use of multiple imaging surveys further mitigates potential biases by reducing the likelihood that artifacts from a single dataset are misidentified as genuine features. In particular, the deeper and higher-resolution HSC-SSP imaging enabled the identification of 
19 additional galaxies in the EGIS sample. However, HSC-SSP coverage is limited, encompassing only 
10.4\% of the EGIS sample. An example of the importance of this additional imaging depth is shown in Figure~\ref{fig:desi_hsc}. In the DESI image, shown in panel (a), no unambiguous tidal structure is visible. In contrast, the deeper HSC-SSP image, shown in panel (b), reveals that a structural feature near the bottom of the galaxy, which may have been dismissed as an artifact or background fluctuation in the DESI data alone, is instead consistent with a faint tidal stream.

Using the additional detections from HSC-SSP and applying a median Wald normal approximation to our complete samples, we estimate that full HSC-SSP coverage would yield an additional $\sim$$147\pm65$ galaxies with tidal features in the EGIS sample. This would increase the corresponding tidal fraction to $\sim$$9.6\pm1.4$\%. Applying this same logic to the EGIPS sample would yield $\sim$$174\pm76$, increasing the fraction to $\sim$$9.4\pm1.4$\%. These results suggest that differences in survey depth and coverage are of great importance for estimating the fraction of LSB features in the local Universe.

The influence of the photometric depth on the detectability of tidal features is further illustrated by the deeper APO images discussed in Section~\ref{sec:apo}, where faint structures are more clearly visible compared to shallower survey data, in which such features are less prominent. \citet{2008ApJ...689..936J} suggest that the majority of tidal features can be detected at average surface brightness levels fainter than $\sim$$30$~mag\,arcsec$^{-2}$. However, reaching such depths is challenging for wide-field surveys, which must balance sky coverage with exposure time and therefore often lack the sensitivity required to detect the faintest structures. 
{Moreover, the limiting surface brightness of an imaging survey is determined not only by its depth but also by the accuracy with which the sky background can be estimated and removed. Accurate sky subtraction remains one of the principal challenges for studies of the LSB Universe, and the optimal strategy depends on both the scientific objectives and the characteristics of the imaging data. Nevertheless, the background subtraction procedures adopted by large deep imaging surveys have several well-recognized limitations. For example, \citet{2026A&C....5501075P} showed that galactic cirrus can lead to systematic sky over-subtraction in deep imaging surveys when it is not accounted for during background estimation. They found that this effect can significantly suppress the measured surface brightness of diffuse structures. Their results demonstrate that galactic cirrus should be explicitly incorporated into sky-background estimation algorithms to avoid the removal of LSB emission. \citet{2024MNRAS.528.4289W} investigated several sky-estimation strategies for LSB imaging using idealized synthetic images. They showed that each approach has distinct limitations: masking and parametric modeling are fundamentally limited by undetected flux, image-combining techniques depend strongly on the observing strategy and introduce additional noise on the smoothing scale, while iterative sky subtraction propagates initial background-estimation errors. They concluded that, for large-scale surveys such as LSST, masking and parametric modeling remain the preferred sky-estimation approach, provided that stellar scattered light is removed before estimating the sky. An important conclusion of these studies is that pipelines designed for LSB science should employ background models that preserve large-scale astrophysical emission, account for galactic cirrus and other spatially varying foregrounds, and avoid estimating the sky on spatial scales comparable to the extent of the target galaxy whenever possible. These considerations are essential for current and future wide-field, deep imaging surveys, where accurate LSB photometry is a primary scientific objective.}

{Given these observational limitations, it is important to quantify the surface-brightness levels that can be reliably reached in practice. Measuring the surface brightness of detected tidal features therefore provides a practical means of assessing the effective observational limits of current imaging datasets.} In this context, \citet{2022A&A...662A.124S} found that tidal structures in their sample typically do not extend beyond a median surface brightness of $\sim$$27.5$~mag\,arcsec$^{-2}$. Motivated by this result, we measured and analyzed the surface brightness distribution of tidal features identified in the EGIS sample.
To perform these measurements, we developed a dedicated graphical user interface (GUI) that enables semi-automated delineation of tidal structures. The GUI allows the user to interactively define polygonal apertures that closely trace the morphology of individual LSB features while avoiding contamination from foreground stars, background galaxies, and unrelated structures. The initial boundaries of the regions can be guided by isophotal contours, after which the regions may be refined manually to better match the observed extent of the tidal debris. The total flux enclosed within each polygon was measured and converted into an average surface brightness by dividing by the aperture area. To ensure the robustness of the measurements, only high-confidence tidal features identified in the EGIS sample were included in the analysis, resulting in 80 galaxies.

\begin{figure}[H]
\includegraphics[width=10cm]{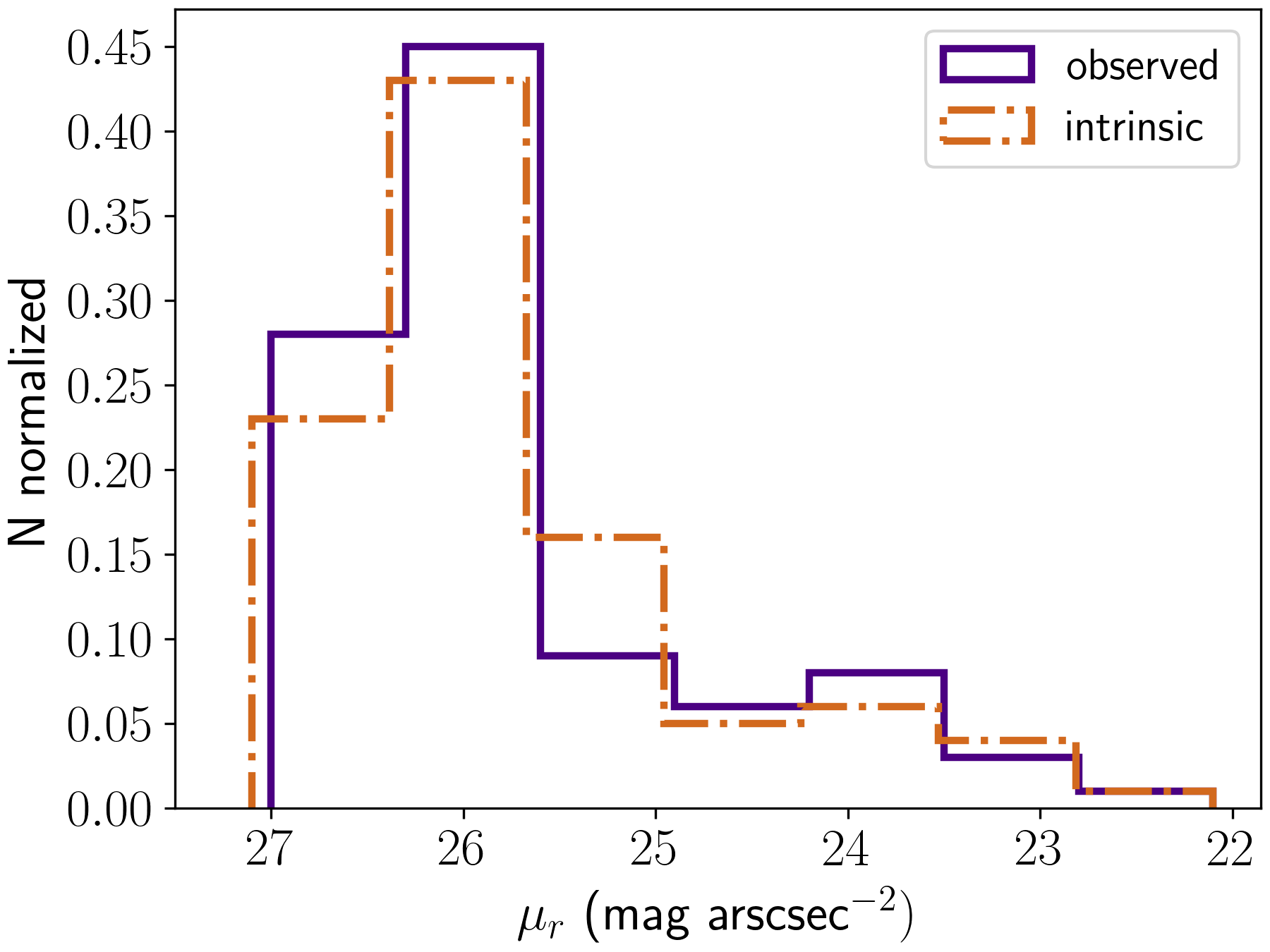}\\
\isPreprints{\centering}{}
\caption{Distribution of the measured median surface brightness in the $r$ band within high-confidence tidal features in the EGIS sample. The distribution is overlaid with the intrinsic surface brightness, corrected for cosmological dimming. A small offset between the observed and intrinsic distributions is evident in both cases.}
  \label{fig:surfbright}
\end{figure}

Our analysis shows that tidal structures are typically detected down to a surface brightness of $\sim$$27.1$~mag\,arcsec$^{-2}$ in the $r$ band (Figure~\ref{fig:surfbright}). The surface-brightness distribution peaks at $25.8 \pm 0.7$~mag\,arcsec$^{-2}$. These values are in good agreement with those reported by \citet{2022A&A...662A.124S}, their mean surface brightness being $25.6 \pm 0.7$~mag\,arcsec$^{-2}$  and with \citet{Stripe82Paper}, whose mean is $26.1 \pm 0.7$~mag\,arcsec$^{-2}$. The similarity of these measurements suggests that the tidal features identified in our sample occupy a comparable surface-brightness regime to those detected in previous deep imaging studies.
The observed distribution likely reflects a combination of physical and observational effects. On the physical side, some of the detected structures may be dynamically young and therefore have not yet fully phase-mixed or faded below the detection threshold. Furthermore, the relatively high stellar masses of the host galaxies may enhance the visibility of tidal debris originating from accretion events. However, observational selection effects are also likely to play an important role. Although the measured surface brightnesses remain well above the nominal photometric limit of the DESI imaging, the detectability of diffuse tidal features depends not only on their mean surface brightness but also on their spatial extent, morphology, and local background fluctuations. As a result, the faintest tidal structures may remain undetected, biasing the observed distribution toward features with higher surface brightness.

\subsection{Redshift-Dependent Detection Biases}

\begin{figure}[H]
\begin{adjustwidth}{-1.5cm}{-0.5cm}
\centering
\subfloat[\centering]{\includegraphics[width=8.5cm]{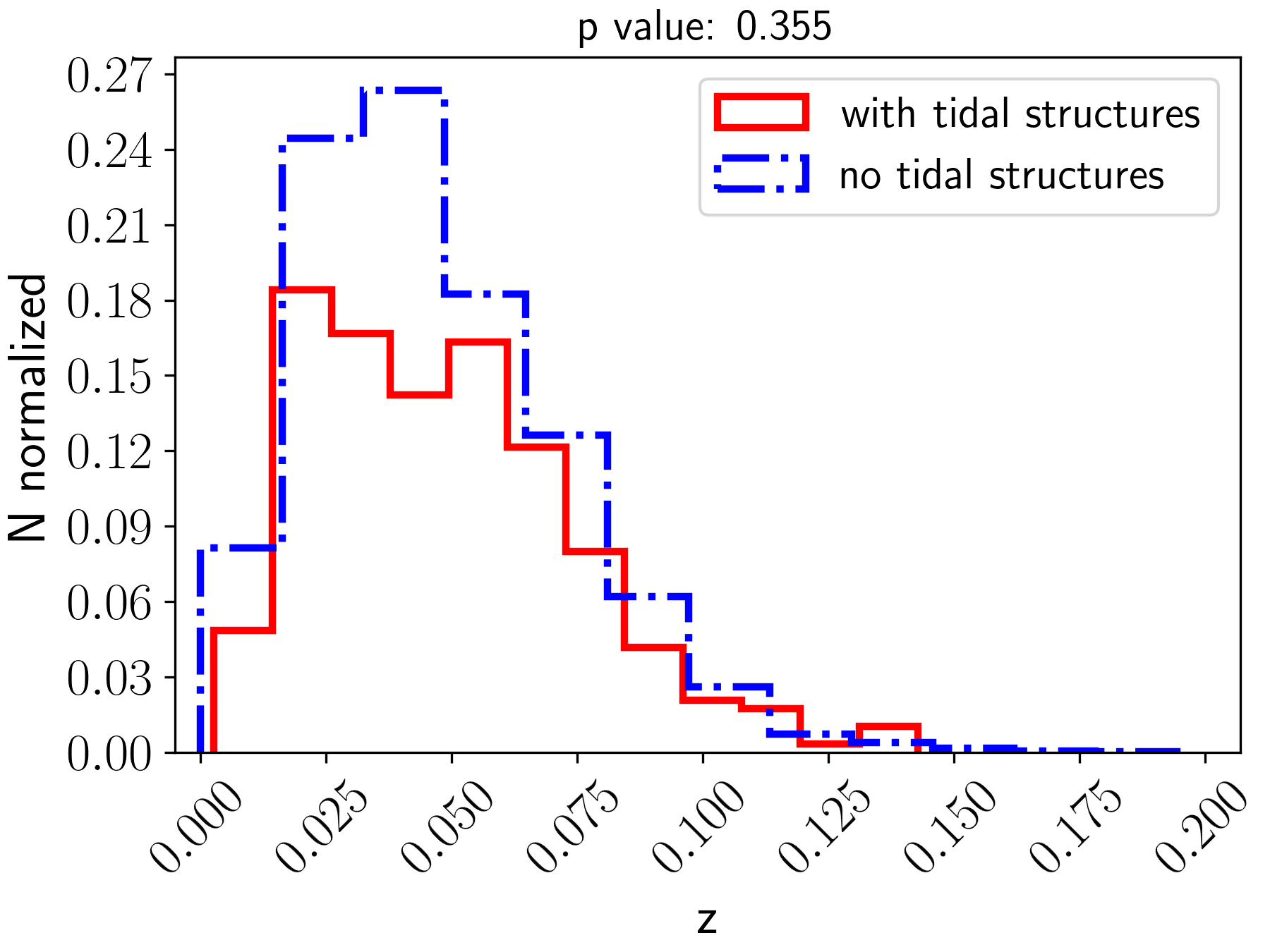}}
\subfloat[\centering]{\includegraphics[width=8.5cm]{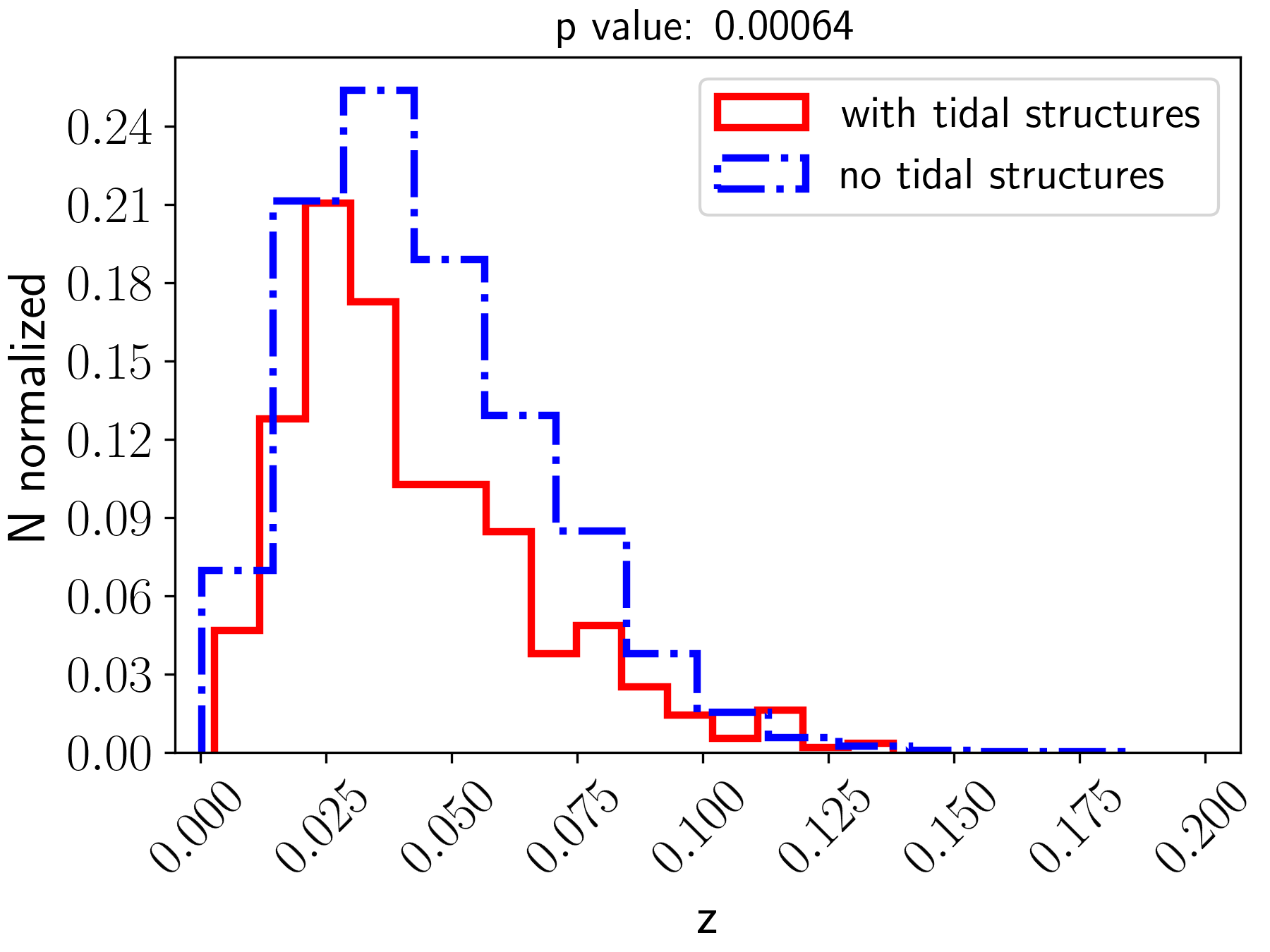}}\\
\end{adjustwidth}
  \caption{Spectroscopic redshift distribution for galaxies from the EGIS (\textbf{a}) and the EGIPS (\textbf{b}) samples.}
  \label{fig:z_spec}
\end{figure}

Figure~\ref{fig:z_spec} shows the spectroscopic redshift distributions of galaxies with and without tidal features for the EGIS and EGIPS samples. For both samples, galaxies hosting tidal features are systematically shifted toward lower redshifts compared to galaxies without such features. The difference in the modal redshift is $\Delta z = 0.03$ for EGIS and $\Delta z = 0.01$ for EGIPS, consistent with the offset of $\Delta z = 0.02$ reported by \citet{Stripe82Paper}.

To assess the statistical significance of this shift, we performed a Mann–Whitney U test, obtaining $p = 0.355$ for EGIS, suggesting that the null hypothesis cannot be rejected. Yet for the EGIPS sample, we obtain $p = 6.4 \times 10^{-4}$, which allows us to reject the null hypothesis that the two populations are drawn from the same parent distribution, indicating that galaxies with tidal features are preferentially found at lower redshifts. This trend is consistent with both observational and simulation-based studies~\citep{2018ApJ...866..103K,Martin2022,2023MNRAS.521.3861D}, which have shown that tidal features become progressively more difficult to detect with increasing redshift owing to cosmological surface-brightness dimming and declining physical spatial resolution. However, the former effect is expected to be negligible for galaxies in our samples. In AB magnitudes, cosmological surface-brightness attenuation scales as \linebreak  $(1+z)^3$~\citep{2020ApJ...903...14W}. For the EGIS and EGIPS samples, we derive mean dimming factors of $1.14 \pm 0.07$ and $1.1 \pm 0.05$, respectively. Such modest attenuation corresponds to only minor changes in surface brightness and is therefore unlikely to significantly affect the detectability of tidal features. This conclusion is further supported by Figure~\ref{fig:surfbright}, which shows that the average difference between observed and intrinsic (pre-dimming) surface brightness values is modest, $\sim$$0.05$ for EGIS.

\subsection{Stellar Mass–Dependent Bias}
\label{sec:mass_bias}

The detectability of tidal features is known to depend on galaxy stellar mass, with more massive galaxies exhibiting a higher incidence of observable structures~\citep{2013ApJ...765...28A,2018ApJ...866..103K,2020MNRAS.498.2138B,Jackson2023}. This trend is expected, as more massive galaxies are generally more luminous, and the tidal debris produced during interactions is correspondingly brighter and easier to detect. In addition, the merger rate increases with stellar mass, further enhancing the likelihood of observable tidal features~\citep{2023MNRAS.519.4920G}. Quantitatively, \citet{2020MNRAS.498.2138B} reported that galaxies with $M_{\star} > 10^{11}~M_{\odot}$ host $\sim$$1.7$ times more tidal structures than lower-mass systems, while \citet{Jackson2023} found a more modest increase of $\sim$$1.2$.

Stellar masses for both the EGIS and EGIPS samples were estimated {using the mid-infrared photometry from the AllWISE Source Catalog~\citep{2019ipac.data...I1W}} obtained with the Wide-field Infrared Survey Explorer (WISE; Wright et al. \cite{2010AJ....140.1868W}), following the calibrations of \citet{Jarrett2023}. First, the absolute W1 magnitude was computed using the luminosity distance derived from the spectroscopic redshift and using a flat $\Lambda$CDM cosmology ($H_0=67.4$ km s$^{-1}$ Mpc$^{-1}$, $\Omega_{\rm M}=0.315$, $\Omega_{\Lambda}=0.685$), and then converted to a W1-band luminosity assuming a solar absolute magnitude of $M_{W1,\odot}=3.24$. For galaxies with {reliable} 
 W1 $-$ W2 colors within the calibrated range ($-0.2 \leq W1-W2 \leq 0.4$ mag), stellar masses were derived using the color-dependent mass-to-light ratio relation of~\citet{Jarrett2023} (see their {Equation (3)).} 
 For galaxies with colors outside this range or lacking reliable W2 photometry, we adopted the luminosity-only calibration based on a third-order polynomial fit between $\log M_{\odot}$ and $\log L_{W1}$ (see Equation (2) in Jarrett et al. \cite{Jarrett2023}).

Our complete galaxy sample that exhibits tidal features has a higher average stellar mass ($\sim$$10^{10.3}~M_{\odot}$) compared to those without tidal features ($\sim$$10^{9.9}~M_{\odot}$) for EGIS. A similar analysis for the EGIPS sample gives the same results ($\sim$$10^{10.2}~M_{\odot}$ and $\sim$$10^{9.9}~M_{\odot}$).

\begin{figure}[H] 
\begin{adjustwidth}{-1.5cm}{-0.5cm}
\centering
\subfloat[\centering]{\includegraphics[width=8.5cm]{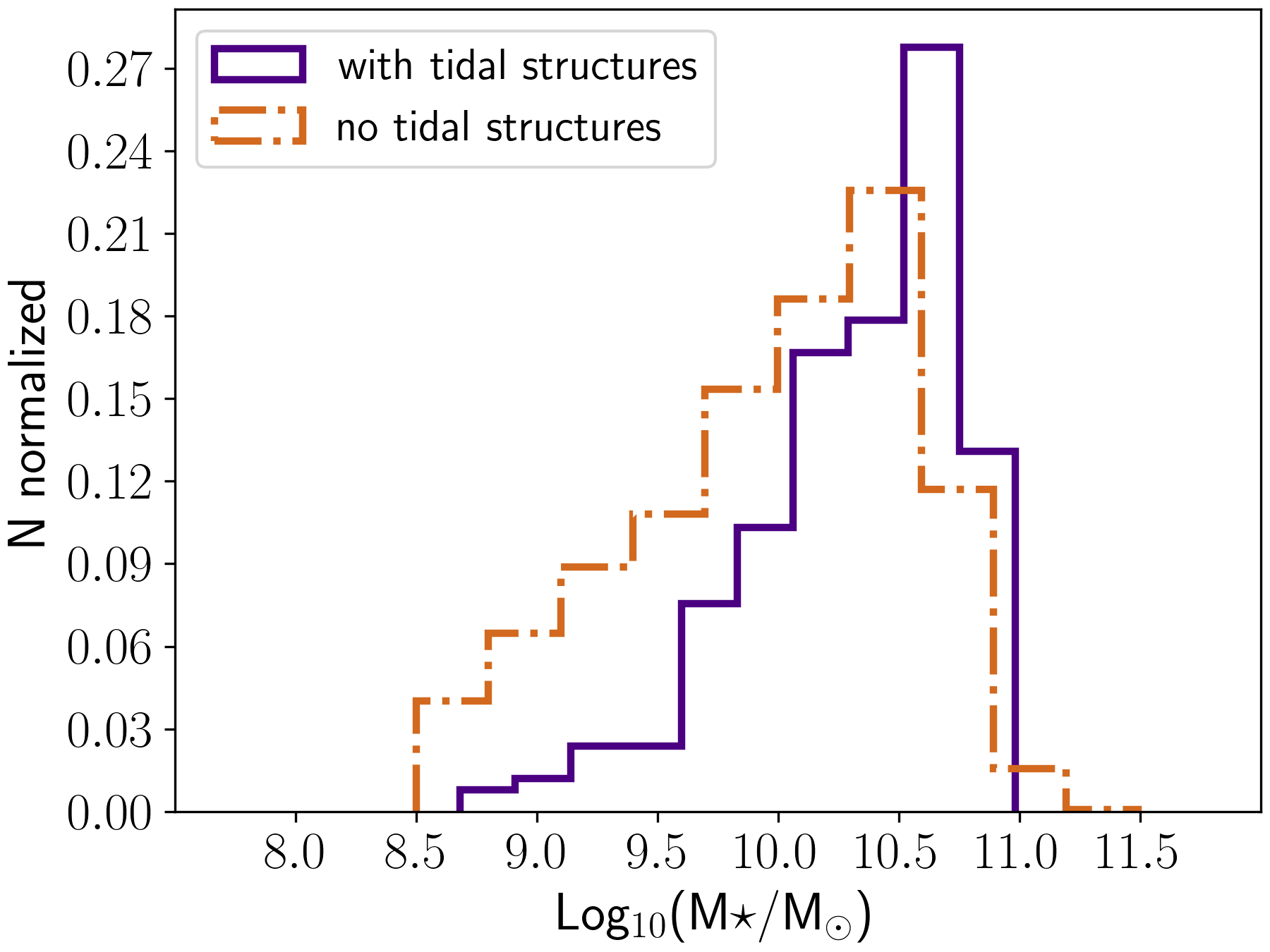}}
\subfloat[\centering]{\includegraphics[width=8.5cm]{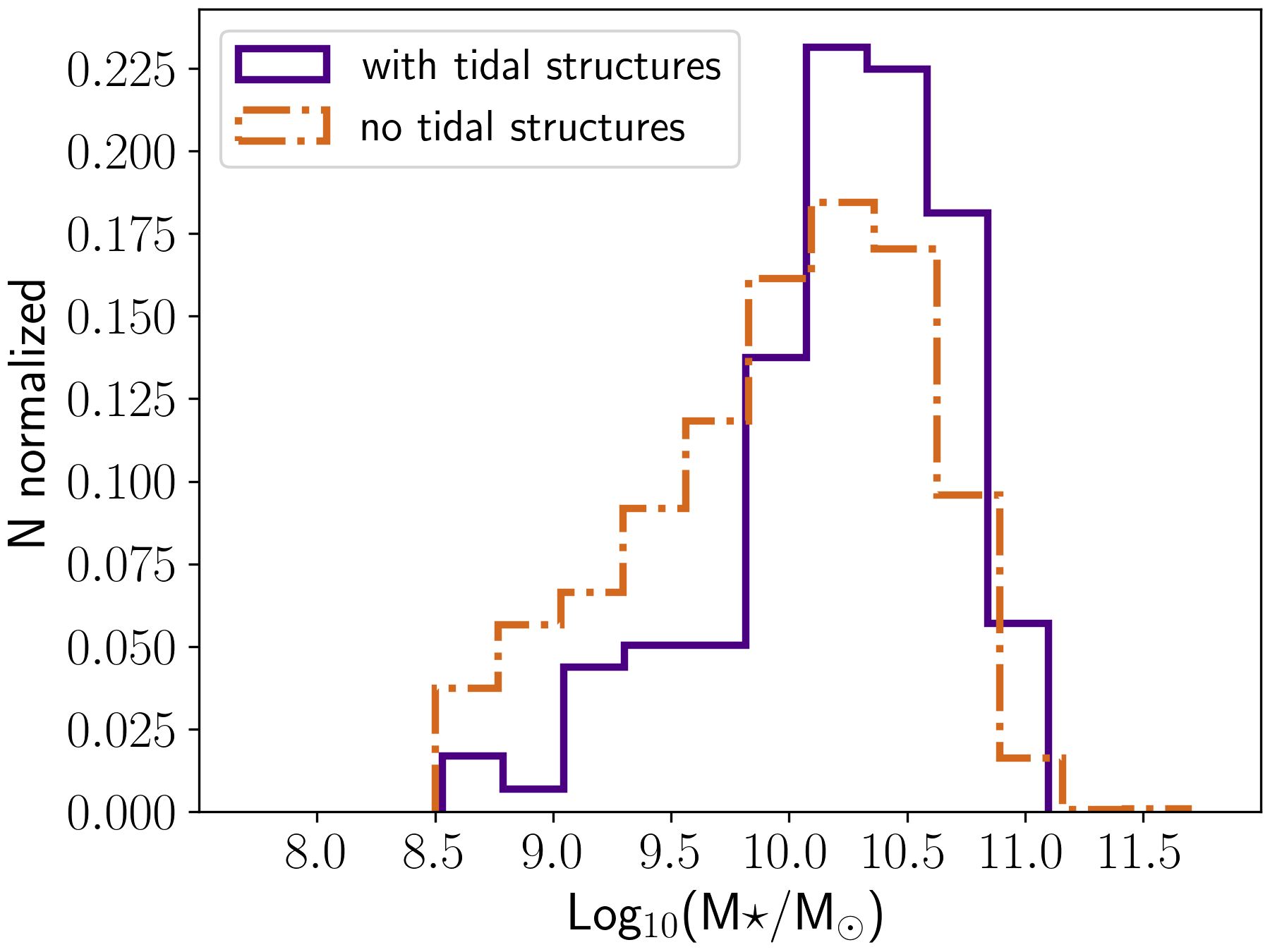}}\\
\end{adjustwidth}
  \caption{{Stellar} 
 mass distribution for galaxies from the EGIS (\textbf{a}) and the EGIPS (\textbf{b}) samples calculated using WISE mid-infrared photometry. Both plots compare the masses of complete galaxies with tidal features and those without.}
  \label{fig:mass}
\end{figure}

As shown in Figure~\ref{fig:mass}, the stellar mass distribution of complete galaxies hosting tidal structures peaks at $\sim$$10^{10.5}~M_{\odot}$ {for both samples}, slightly lower than what is stated {in} previous studies~\citep{Jackson2023,Stripe82Paper}. This offset may be influenced by the incomplete availability of stellar mass estimates for the {EGIS and EGIPS galaxies}, which could bias the inferred distribution. For each catalog, we performed a Mann--Whitney U test comparing the stellar mass distributions of galaxies with tidal structures to those without. The test was applied independently to the EGIS and EGIPS samples. We obtain $p = 1\times10^{-23}$ for EGIS and $p = 9.6\times10^{-16}$ for EGIPS, allowing us to reject the null hypothesis and indicating that the two sub-samples for each sample are drawn from statistically distinct populations.

\begin{figure}[H]
\includegraphics[width=11cm]{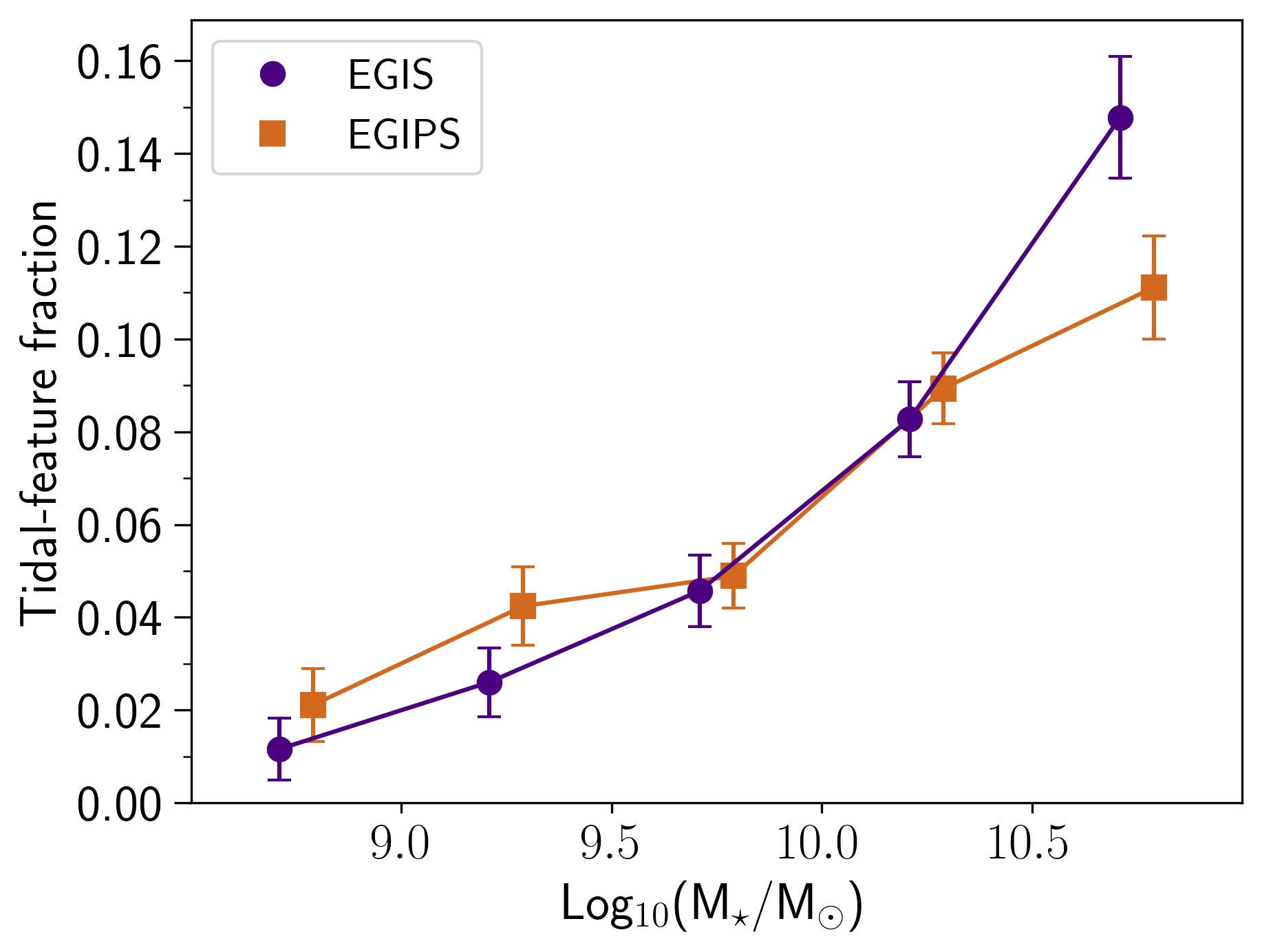}\\
\isPreprints{\centering}{}
\caption{{Tidal fraction for galaxies divided into stellar mass bins for the complete EGIS and EGIPS samples. The error bars represent the $1\sigma$ binomial uncertainties. The points are offset slightly to prevent overlap.}}
  \label{fig:tidal_vs_mass}
\end{figure}

Overall, Figure~\ref{fig:mass} confirms a positive correlation between stellar mass and the presence of tidal structures in our sample. {To further illustrate this relationship, we calculated the tidal feature fraction in 0.5-dex stellar mass bins for EGIS and EGIPS galaxies with available stellar mass estimates, as seen in Figure~\ref{fig:tidal_vs_mass}. Consistent with \citet{Sola2025}, the tidal fraction increases with stellar mass, reaching 11\% and 15\% for EGIPS and EGIS respectively for masses Log$_{10}(M_{\star}/M_{\odot}) > 10.5$.}

\subsection{Comparison to Observational Studies}
\label{sec:comp_obs}

To place our results in a broader observational context, we compare the incidence of LSB features in our sample with those reported in previous studies based on deep imaging surveys. Such comparisons should be interpreted with caution, as the samples differ in stellar mass distributions, redshift ranges, galaxy populations, and methodologies used to estimate limiting surface-brightness depths. We do not attempt to homogenize these measurements, since a consistent recalculation of the surface-brightness limits for all datasets is beyond the scope of this work. Instead, we focus on studies reaching depths broadly comparable to those of our survey. For consistency, we use the complete EGIS and EGIPS samples defined in Section~\ref{sec:sample}.

Among the studies considered here, the closest agreement with our results is found by \citet{2018ApJ...866..103K}, who analyzed 21{,}208 galaxies from the HSC-SSP survey reaching a limiting surface brightness of $28.1$~mag\,arcsec$^{-2}$ in the $r$ band. Tidal features were identified using a CNN and subsequently verified through visual inspection. Of the galaxies in their sample, 1201 (5.6\%) were found to host at least one tidal feature. This fraction is remarkably similar to that obtained for our complete EGIS and EGIPS samples. However, it remains unclear whether their sample is complete with respect to the selection criteria adopted in that study, which complicates a direct comparison.

Several other studies report substantially higher fractions of galaxies hosting LSB features. For example, \citet{2013ApJ...765...28A} visually inspected 1781 luminous galaxies ($M_r<-19.3$) in the Canada--France--Hawaii Telescope Legacy Survey (CFHTLS) over the redshift range $0.04<z<0.2$. Their classification scheme included streams, shells, fans, linear features, and diffuse structures. Restricting the sample to the highest-confidence detections, they found that approximately 12\% of galaxies exhibit tidal features, with the fraction increasing to 18\% when weaker but convincing features were included and to 26\% when marginal detections were considered. The higher incidence is likely related to the strong bias of their sample toward luminous and massive galaxies, with tidal features being particularly common in galaxies with stellar masses above $10^{10.5}$~M$_\odot$ and in red galaxies. In contrast, the EGIS and EGIPS samples span a broader range of stellar masses and include a larger fraction of lower-mass galaxies, for which the incidence of tidal features is known to be lower (Section~\ref{sec:mass_bias}).

A similarly elevated fraction was reported by \citet{2018ApJ...857..144H}, who investigated 1048 galaxies from the nearby volume-limited RESOLVE survey using DECaLS imaging with an $r$-band surface-brightness limit of approximately 27.9~mag\,arcsec$^{-2}$. After masking foreground sources and applying image smoothing, they identified tidal features in $17\pm2$\% of the galaxies. Compared to our samples, the RESOLVE galaxies are generally closer, larger in angular size, and therefore better resolved. In addition, \citet{2018ApJ...857..144H} found that tidal features are preferentially associated with gas-rich galaxies and, among gas-poor galaxies, with higher stellar masses. Both effects likely contribute to the higher fraction observed in their study.

Likewise, \citet{2018A&A...614A.143M} analyzed a mass-complete sample of 297 galaxies from the Spitzer Survey of Stellar Structure in Galaxies (S$^{4}$G, Sheth et al. \cite{2010PASP..122.1397S}), complete above a stellar mass threshold of $10^{9.2}$~M$_\odot$. Using SDSS imaging, they reported tidal feature fractions ranging from 9\% to 14\%, depending on the adopted confidence threshold. Although these values exceed those found in our samples, the difference is relatively modest and can plausibly be attributed to the greater representation of massive galaxies in their sample.

\textls[-15]{The importance of stellar mass becomes even more apparent in studies specifically targeting massive galaxies. Studies targeting the most massive galaxies in dense environments report some of the highest incidences of LSB features. For example,~\citet{2020ApJS..247...43K} found tidal streams in 22\% and shells in 9.4\% of 170 Brightest Cluster Galaxies. Such galaxies occupy the centers of massive clusters, where repeated mergers, satellite accretion, and tidal stripping are expected to be common. Consequently, their elevated LSB feature fractions are consistent with their unique environments and assembly histories rather than indicative of a discrepancy with our results.}

Using deep imaging from the MATLAS survey and CFIS, \citet{2022A&A...662A.124S} examined 352 nearby galaxies within 42~Mpc and found tidal features in 127 objects (36\%). Their sample was designed to target massive nearby galaxies, with a strong emphasis on early-type galaxies. The combination of high stellar masses, low redshifts, and deeper imaging naturally enhances the detectability of LSB features compared to the more distant and morphologically diverse galaxies in our samples.

An even higher fraction was reported by \citet{Jackson2023}, who investigated the LSB outskirts of 118 massive central galaxies using HSC-SSP imaging. Approximately 42\% of their galaxies exhibit tidal features, while an additional 43\% display diffuse stellar haloes. Since their sample consists exclusively of massive central galaxies selected to have assembled most of their stellar mass at early epochs, such a high incidence of LSB features is not unexpected.

A complementary perspective is provided by \citet{Sola2025}, who visually inspected 19{,}387 nearby galaxies ($z\leq0.02$) from the Siena Galaxy Atlas 2020~\citep{2023ApJS..269....3M} using original, model, and residual DESI images. They found that $11.9\pm0.2$\% of galaxies host tidal features. Although this fraction remains higher than that measured in our study, the difference is readily explained by their focus on very nearby galaxies, whose larger angular sizes facilitate the detection of faint LSB structures. Moreover, their diameter-selected sample favors large, well-resolved galaxies, and their extensive use of residual images increases sensitivity to weak tidal signatures. They also report a strong dependence of tidal feature incidence on stellar mass, with the fraction increasing from $2.4\pm0.4$\% at $\sim$$10^{8}$~M$_\odot$ to $36.5\pm1.2$\% at $\sim$$5\times10^{11}$~M$_\odot$, further supporting the importance of mass selection effects.

Overall, the incidence of LSB features reported in the literature spans a wide range, from $\sim$$5$\% to more than 40\%, even among studies reaching comparable surface-brightness depths. Among the surveys considered here, the HSC-SSP study of \citet{2018ApJ...866..103K} yields a tidal feature fraction of 5.6\%, in excellent agreement with the values derived for our complete EGIS and EGIPS samples. In contrast, studies targeting nearby, luminous, or massive galaxies generally report substantially higher fractions. Examples include the CFHTLS sample of \citet{2013ApJ...765...28A}, the RESOLVE survey~\citep{2018ApJ...857..144H}, the S$^{4}$G sample analyzed by \citet{2018A&A...614A.143M}, the MATLAS/CFIS study of \citet{2022A&A...662A.124S}, and the sample of massive central galaxies investigated by \citet{Jackson2023}. The highest fractions are found in studies focusing on exceptionally massive galaxies, such as BCGs~\citep{2020ApJS..247...43K} and nearby massive central galaxies~\citep{Jackson2023}, where hierarchical assembly and repeated accretion events are expected to be particularly important.

A consistent picture emerges from these comparisons. The observed fraction of galaxies hosting LSB features depends not only on imaging depth, but also on galaxy stellar mass, redshift, angular size, environment, and the adopted feature identification methodology. In particular, the incidence of LSB features increases strongly with stellar mass, while lower redshifts and larger angular sizes improve the detectability of diffuse structures. The use of residual images, image smoothing, or specialized detection tools can further enhance sensitivity to faint tidal signatures. Consequently, differences in reported LSB feature fractions are likely driven by a combination of physical and observational effects rather than by genuine inconsistencies between studies.

Taken together, the available evidence suggests that the $\sim$$6$\% incidence of LSB features measured in our complete EGIS and EGIPS samples is consistent with expectations for galaxy populations spanning a broad range of stellar masses and environments at a mean redshift of $z\sim0.05$. The higher fractions reported in many previous studies can largely be explained by their focus on nearby, massive, or otherwise preferentially selected galaxies, for which both the intrinsic occurrence and the detectability of LSB features are enhanced.

\subsection{Comparison to Simulations}
\label{sec:comp_sim}

Having compared our results with previous observational studies, it is also instructive to examine expectations from numerical simulations. In contrast to the observational studies discussed above, which report LSB feature fractions spanning a wide range depending on sample selection and methodology, simulations provide a means of assessing the intrinsic frequency of tidal structures and their dependence on observational depth. As shown below, simulation-based predictions have likewise evolved considerably over time, reflecting improvements in numerical resolution, physical modeling, and the generation of increasingly realistic mock observations.

One of the earliest attempts to quantify the detectability of tidal debris was carried out by \citet{2001ApJ...557..137J}, who combined semi-analytic prescriptions with simplified $N$-body simulations to investigate the visibility of tidal features at different surface-brightness limits. Although the details of the simulations were not reported, their analysis of approximately 100 host galaxies suggested that, at a limiting surface brightness of $\sim$$29$~mag\,arcsec$^{-2}$, roughly 20--40\% of galaxies should exhibit detectable tidal features. These values are substantially higher than the fractions measured in our study and in several recent observational surveys. In hindsight, such predictions may partly reflect the limitations of early simulations, including lower numerical resolution and less sophisticated treatments of baryonic physics and galaxy evolution~\citep{2024A&A...683A.181B}.

More recent cosmological simulations have enabled substantially more realistic investigations of tidal feature detectability. For example, \citet{Martin2022} used mock observations generated from the NEWHORIZONS cosmological simulation~\citep{2021A&A...651A.109D} to assess the visibility of tidal features in the context of LSST observations. Their analysis demonstrated that the fraction of galaxies with detectable tidal features depends strongly on both limiting surface brightness and redshift, with deeper imaging recovering progressively fainter structures. Based on \citet{Martin2022} {Figure~17} 
 and adopting the mean redshift of our EGIS and EGIPS samples, a limiting surface brightness of $\sim$$28$~mag\,arcsec$^{-2}$ in the $r$ band corresponds to an expected tidal feature fraction of approximately 30\%. This prediction remains significantly higher than the fractions measured in our observational samples.

Similarly, \citet{2024A&A...686A.182V} used the Magneticum Pathfinder simulation to investigate the occurrence of tidal tails, streams, and shells as a function of galaxy morphology. Consistent with observational studies, they found that tidal features are more common in early-type galaxies than in late-type galaxies. Their sample of 520 mock galaxies with stellar masses $M_{\star} \ge 2.4 \times 10^{10}~M_{\odot}$ yielded a tidal feature fraction of approximately 21\%, somewhat lower than the prediction of \citet{Martin2022} but still substantially above the values derived from our EGIS and EGIPS samples. More broadly, recent simulation-based studies generally predict tidal feature fractions in the range of $\sim$$20$--40\%, including \linebreak  $\sim$$32$--40\% from \citet{2024MNRAS.530.4422K} and $\sim$$18$--30\% from \citet{2022MNRAS.514.4898V}. Taken together, these results suggest a persistent tension between many simulation-based predictions and observational measurements.

However, this picture has begun to change with the advent of higher-resolution cosmological simulations and more realistic mock-observing pipelines. In particular, \citet{2025A&A...700A.176M} reported tidal feature fractions much closer to those measured observationally. Using mock images generated from the Copernicus Complexio (COCO; Hellwing et al. \cite{2016MNRAS.457.3492H}), IllustrisTNG TNG50~\citep{2019MNRAS.490.3234N}, and Auriga~\citep{2017MNRAS.467..179G} simulations, they examined the detectability of tidal features over a range of $r$-band surface-brightness limits. For TNG50, they found a tidal feature fraction of approximately 7\% at a limiting surface brightness of 28.6~mag\,arcsec$^{-2}$, comparable to the depth of our observations and in good agreement with both our results and those of \citet{Stripe82Paper}. While the TNG50 and Auriga simulations broadly reproduce the observed incidence of tidal features, COCO systematically overpredicts their abundance. The authors attribute this discrepancy to the particle-tagging technique employed in COCO, which produces tidal streams and shells that are both more numerous and longer-lived than those found in hydrodynamical simulations.

The contrast between these studies highlights the importance of both numerical resolution and physical realism in predicting the abundance of tidal features. Modern cosmological simulations incorporate increasingly sophisticated treatments of baryonic processes, stellar evolution, feedback from stars and active galactic nuclei, and the coupled evolution of dark matter and baryons. These improvements appear to reduce the prevalence of long-lived, easily detectable tidal debris compared to earlier models. The close agreement between the TNG50 and Auriga predictions and observational measurements supports the view that realistic galaxy formation physics, combined with sufficiently high resolution, is essential for reproducing the observed incidence of LSB features.

Overall, the comparison with simulations suggests that the apparent discrepancy between theoretical predictions and observations has narrowed considerably in recent years. While many earlier studies predicted that 20--40\% of galaxies should host detectable tidal features, the latest generation of high-resolution cosmological simulations yields fractions closer to the $\sim$$6$--7\% measured in our complete EGIS and EGIPS samples. This convergence indicates that improvements in both observational realism and physical modeling are crucial for obtaining reliable predictions of the abundance of LSB features.

\section{Summary and Conclusions}
\label{sec:conclusions}

In this paper, we analyzed two large, statistically well-defined samples of edge-on galaxies drawn from the EGIS (5606 galaxies) and EGIPS (14{,}237 galaxies) catalogs. The EGIS sample combines imaging from DESI and HSC-SSP and is further complemented by targeted deep follow-up observations of 14 galaxies obtained with the ARCTIC instrument on the APO 3.5-m telescope. Tidal features were identified through visual inspection and classified according to established morphological categories, including tails and streams, shells, plumes, fans, asymmetric stellar halos, bridges, arcs, loops, and tidally disrupted satellites. Particular attention was paid to the identification and mitigation of contamination by galactic cirrus. The occurrence and properties of these features were then quantified for both the EGIS and EGIPS samples.

The main conclusions of this study are summarized as follows:

\begin{enumerate}
\item The multi-survey datasets used in this study enabled a robust classification of LSB tidal structures. The EGIS and EGIPS samples are particularly well suited for the detection of such features, as they consist predominantly of nearby galaxies that are only weakly affected by cosmological surface-brightness dimming and are dominated by relatively massive systems, in which tidal debris is expected to be more readily detectable. Furthermore, the edge-on orientation of the galaxies reduces projection-related ambiguities, while the depth of the DESI imaging is sufficient to recover tidal features with surface brightnesses comparable to those reported in previous studies.

\item The combination of imaging from several surveys and dedicated image-processing techniques improves the reliability of LSB feature identification and reduces the risk of misclassifying imaging artifacts, foreground contamination, or other spurious structures as genuine tidal features.

\item Deep APO imaging demonstrates that many LSB features identified in DESI data are components of more extended and morphologically complex structures {that} cannot be inferred from shallower imaging alone. The APO observations reveal previously undetected tidal debris in several galaxies and clarify the extent of structures already visible in DESI, underscoring the importance of deep imaging for reconstructing the recent accretion and interaction histories of nearby galaxies.

\item In the complete EGIS subsample, 295 galaxies (6.4\%) exhibit tidal structures, while the complete EGIPS subsample contains 338 such galaxies (6.2\%). The close agreement between the two independent samples indicates that the incidence of detectable tidal features at a typical surface-brightness depth of $\mu_r \approx 28.6$~mag\,arcsec$^{-2}$ is $\sim$$6$\%. This value is consistent with previous observational studies of galaxies spanning similar ranges of stellar mass, environment, and redshift (see the discussion in Section~\ref{sec:comp_obs}).

\item {Using galaxies from the statistically complete EGIS and EGIPS subsamples that have available stellar mass estimates, we find a clear increase in tidal feature incidence with stellar mass. For galaxies with $\log_{10}(M_{\star}/M_{\odot})>10.5$, the tidal feature fraction reaches 15\% and 11\% for EGIS and EGIPS respectively. The consistency of this trend between the two independent samples supports a genuine relationship between stellar mass and the detectability or occurrence of tidal structures.}

\item Comparison with cosmological simulations indicates that the discrepancy between predicted and observed frequencies of tidal features has decreased substantially (see Section~\ref{sec:comp_sim}). While many earlier simulations predicted detectable tidal structures in 20--40\% of galaxies, recent high-resolution simulations produce fractions much closer to the $\sim$$6$\% measured in the EGIS and EGIPS samples, highlighting the importance of numerical resolution, realistic mock observations, and improved galaxy formation models.
\end{enumerate}

Looking ahead, continued improvements in both simulations and observations are expected to reduce the current discrepancies. As simulations become more sophisticated and observational data reach greater depths, the comparison between theory and observation will become increasingly robust. In particular, forthcoming deep, wide-field surveys such as the LSST will provide large, statistically complete samples, enabling more precise measurements of tidal feature incidence in the local Universe and facilitating more meaningful comparisons with cosmological simulations.

\vspace{6pt} 





\authorcontributions{K.R.A. led the data preparation, analyzed the data, double-checked classifications, and wrote most of this manuscript while under the supervision and guidance of A.M. Conceptualization: A.M. and K.R.A.;  Project Administration: A.M. and K.R.A.; Methodology: A.M. and K.R.A.; Software (Code Contribution): K.R.A., A.M., J.S., J.G., L.S., and T.S.; Data Curation: K.R.A., A.M., J.S., L.S., and T.S.; Validation: K.R.A., A.M., L.S., and T.S.; Formal Analysis: K.R.A., A.M., J.S., L.S., and T.S.; Writing---original draft preparation: K.R.A., A.M., and J.S.; Writing---review and editing: K.R.A. and A.M. All authors have read and agreed to the published version of the manuscript.}

\funding{This research received no external funding} 

\dataavailability{The full image catalog of LSB tidal structures and corresponding data can be found at \url{https://doi.org/10.5281/zenodo.20573582}. The original EGIS data can be found at \url{http://users.apo.nmsu.edu/~dmbiz/EGIS/}  ({accessed on 1 January 2024}). The IMAN code can be found at \url{https://bitbucket.org/mosenkov/iman_new/src/master/} ({accessed on 1 January 2025}). {The IRAS maps can be found at \url{https://lambda.gsfc.nasa.gov/product/iras/iras_iris_get.html} ({accessed on 1 September 2022}).}}

\acknowledgments{{We} 
 thank the anonymous {reviewers} for their thorough and constructive referee's reports, which helped to improve the paper. {Based on observations obtained with Apache Point Observatory 3.5-m Astrophysical Research Consortium Telescope.} This research has made use of the NASA/IPAC Infrared Science Archive (IRSA; \url{http://irsa.ipac.caltech.edu/frontpage/} {accessed on 1 September 2022}), and the NASA/IPAC Extragalactic Database (NED; \url{https://ned.ipac.caltech.edu/} {accessed on 1 January 2024}), both of which are operated by the Jet Propulsion Laboratory, California Institute of Technology, under contract with the National Aeronautics and Space Administration. {We acknowledge the use of the Legacy Archive for Microwave Background Data Analysis (LAMBDA; \url{https://lambda.gsfc.nasa.gov/} {accessed on 1 September 2022}), part of the High Energy Astrophysics Science Archive Center (HEASARC). HEASARC/LAMBDA is a service of the Astrophysics Science Division at the NASA Goddard Space Flight Center.} This research has made use of the HyperLeda database (\url{http://leda.univ-lyon1.fr/} {accessed on 1 January 2024;} \cite{2014A&A...570A..13M}). Based on observations from the DESI Legacy Surveys. The Legacy Surveys consist of three individual and complementary projects: the Dark Energy Camera Legacy Survey (DECaLS; Proposal ID \#2014B-0404; PIs: David Schlegel and Arjun Dey), the Beijing-Arizona Sky Survey (BASS; NOAO Prop. ID \#2015A-0801; PIs: Zhou Xu and Xiaohui Fan), and the Mayall z-band Legacy Survey (MzLS; Prop. ID \#2016A-0453; PI: Arjun Dey). DECaLS, BASS and MzLS together include data obtained, respectively, at the Blanco telescope, Cerro Tololo Inter-American Observatory, NSF’s NOIRLab; the Bok telescope, Steward Observatory, University of Arizona; and the Mayall telescope, Kitt Peak National Observatory, NOIRLab. Pipeline processing and analyses of the data were supported by NOIRLab and the Lawrence Berkeley National Laboratory (LBNL). The Legacy Surveys project is honored to be permitted to conduct astronomical research on Iolkam Du’ag (Kitt Peak), a mountain with particular significance to the Tohono O’odham Nation. NOIRLab is operated by the Association of Universities for Research in Astronomy (AURA) under a cooperative agreement with the National Science Foundation. LBNL is managed by the Regents of the University of California under contract to the U.S. Department of Energy.
This project used data obtained with the Dark Energy Camera (DECam), which was constructed by the Dark Energy Survey (DES) collaboration. Funding for the DES Projects has been provided by the U.S. Department of Energy, the U.S. National Science Foundation, the Ministry of Science and Education of Spain, the Science and Technology Facilities Council of the United Kingdom, the Higher Education Funding Council for England, the National Center for Supercomputing Applications at the University of Illinois at Urbana-Champaign, the Kavli Institute of Cosmological Physics at the University of Chicago, Center for Cosmology and Astro-Particle Physics at the Ohio State University, the Mitchell Institute for Fundamental Physics and Astronomy at Texas A\&M University, Financiadora de Estudos e Projetos, Fundacao Carlos Chagas Filho de Amparo, Financiadora de Estudos e Projetos, Fundacao Carlos Chagas Filho de Amparo a Pesquisa do Estado do Rio de Janeiro, Conselho Nacional de Desenvolvimento Cientifico e Tecnologico and the Ministerio da Ciencia, Tecnologia e Inovacao, the Deutsche Forschungsgemeinschaft and the Collaborating Institutions in the Dark Energy Survey. The Collaborating Institutions are Argonne National Laboratory, the University of California at Santa Cruz, the University of Cambridge, Centro de Investigaciones Energeticas, Medioambientales y Tecnologicas-Madrid, the University of Chicago, University College London, the DES-Brazil Consortium, the University of Edinburgh, the Eidgenossische Technische Hochschule (ETH) Zurich, Fermi National Accelerator Laboratory, the University of Illinois at Urbana-Champaign, the Institut de Ciencies de l’Espai (IEEC/CSIC), the Institut de Fisica d’Altes Energies, Lawrence Berkeley National Laboratory, the Ludwig Maximilians Universitat Munchen and the associated Excellence Cluster Universe, the University of Michigan, NSF’s NOIRLab, the University of Nottingham, the Ohio State University, the University of Pennsylvania, the University of Portsmouth, SLAC National Accelerator Laboratory, Stanford University, the University of Sussex, and Texas A\&M University. BASS is a key project of the Telescope Access Program (TAP), which has been funded by the National Astronomical Observatories of China, the Chinese Academy of Sciences (the Strategic Priority Research Program “The Emergence of Cosmological Structures” Grant \# XDB09000000), and the Special Fund for Astronomy from the Ministry of Finance. The BASS is also supported by the External Cooperation Program of Chinese Academy of Sciences (Grant \# 114A11KYSB20160057), and Chinese National Natural Science Foundation (Grant \# 12120101003, \# 11433005).
The Legacy Survey team makes use of data products from the Near-Earth Object Wide-field Infrared Survey Explorer (NEOWISE), which is a project of the Jet Propulsion Laboratory/California Institute of Technology. NEOWISE is funded by the National Aeronautics and Space Administration.
The Legacy Surveys imaging of the DESI footprint is supported {by the Director, Office of Science,} 
 Office of High Energy Physics of the U.S. Department of Energy under Contract No. DE-AC02-05CH1123, by the National Energy Research Scientific Computing Center, a DOE Office of Science User Facility under the same contract; and by the U.S. National Science Foundation, Division of Astronomical Sciences under Contract No. AST-0950945 to NOAO. 
Based in part on data collected at the Subaru Telescope and retrieved from the HSC data archive system, which is operated by Subaru Telescope and Astronomy Data Center at National Astronomical Observatory of Japan. The Hyper Suprime-Cam (HSC) collaboration includes the astronomical communities of Japan and Taiwan, and Princeton University. The HSC instrumentation and software were developed by the National Astronomical Observatory of Japan (NAOJ), the Kavli Institute for the Physics and Mathematics of the Universe (Kavli IPMU), the University of Tokyo, the High Energy Accelerator Research Organization (KEK), the Academia Sinica Institute for Astronomy and Astrophysics in Taiwan (ASIAA), and Princeton University. Funding was contributed by the FIRST program from Japanese Cabinet Office, the Ministry of Education, Culture, Sports, Science and Technology (MEXT), the Japan Society for the Promotion of Science (JSPS), Japan Science and Technology Agency (JST), the Toray Science Foundation, NAOJ, Kavli IPMU, KEK, ASIAA, and Princeton University. This paper makes use of software developed for the Large Synoptic Survey Telescope. We thank the LSST Project for making their code available as free software at  \url{http://dm.lsst.org} {(accessed on 1 April 2025).}
The Pan-STARRS1 Surveys (PS1) have been made possible through contributions of the Institute for Astronomy, the University of Hawaii, the Pan-STARRS Project Office, the Max-Planck Society and its participating institutes, the Max Planck Institute for Astronomy, Heidelberg and the Max Planck Institute for Extraterrestrial Physics, Garching, The Johns Hopkins University, Durham University, the University of Edinburgh, Queen’s University Belfast, the Harvard-Smithsonian Center for Astrophysics, the Las Cumbres Observatory Global Telescope Network Incorporated, the National Central University of Taiwan, the Space Telescope Science Institute, the National Aeronautics and Space Administration under Grant No. NNX08AR22G issued through the Planetary Science Division of the NASA Science Mission Directorate, the National Science Foundation under Grant No. AST-1238877, the University of Maryland, and Eotvos Lorand University (ELTE) and the Los Alamos National Laboratory. This publication makes use of data products from the Wide-field Infrared Survey Explorer, which is a joint project of the University of California, Los Angeles, and the Jet Propulsion Laboratory/California Institute of Technology, funded by the National Aeronautics and Space Administration. {This work made use of Astropy {6.1.2} (\url{https://www.astropy.org} {accessed on 1 September 2022}) a community-developed core {Python 3.10.12} 
 package and an ecosystem of tools and resources for astronomy~\citep{2013A&A...558A..33A, 2018AJ....156..123A, 2022ApJ...935..167A}. This research made use of Photutils 1.13.0, an {Astropy 6.1.2} package for
detection and photometry of astronomical sources~\citep{2018zndo...1340699B}.}

}

\conflictsofinterest{The authors declare no conflicts of interest. The funders had no role in the design of the study; in the collection, analyses, or interpretation of data; in the writing of the manuscript; or in the decision to publish the results.} 



\abbreviations{Abbreviations}{
The following abbreviations are used in this manuscript:

\noindent 
\begin{longtable}{@{}ll}
EGIS & Edge-on Galaxies In SDSS\\
EGIPS & Edge-on Galaxies in the Pan-STARRS survey\\
{EGIDE} & {The Edge-on Galaxies in the DESI survey}\\
{Pan-STARRS} & {Panoramic Survey Telescope and Rapid Response System}\\
{FITS} & {Flexible Image Transport System}\\
SDSS & Sloan Digital Sky Survey\\
DESI & Dark Energy Spectroscopic Instrument\\
HSC-SSP & Hyper Suprime-Cam Subaru Strategic Program\\
APO & Apache Point Observatory\\
{ARCTIC} & {Astrophysical Research Consortium Telescope Imaging Camera}\\
{ARC} & {Astrophysical Research Consortium}\\
{NASA} & {National Aeronautics and Space Administration}\\
{HyperLeda} & {Hyper-linked Lyon-Meudon Extragalactic Database}\\
{GUI} & {Graphical User Interface}\\
LSST & Legacy Survey of Space and Time\\
LSB & Low surface brightness\\
ETG & Early-type galaxy\\
IMAN & IMage ANalysis\\
DR & Data Release\\
PDR & Public Data Release\\
RGB & Red, Green, Blue\\
HI & Atomic Hydrogen\\
IR & Infrared\\
PSF & Point Spread Function\\
FWHM & Full Width at Half Maximum\\
{CNN} & {Convolutional Neural Network}\\
{CFHTLS} & {Canada--France--Hawaii Telescope Legacy Survey}\\
{RESOLVE} & {Resolved Spectroscopy Of a Local Volume}\\
{WISE} & {Wide-field Infrared Survey Explorer}\\
{MATLAS} & {Mass Assembly of early-Type Galaxies with their fine Structures}\\
{CFIS} & {Canada--France Imaging Survey}\\
{S$^{4}$G} & {Spitzer Survey of Stellar Structure in Galaxies}\\
{BCG} & {Brightest Cluster Galaxy}\\
TNG & The Next Generation\\
COCO & Copernicus Complexio\\
SKIRT & Simulator for Knowledge on Interstellar Radiation and Transfer\\
$\Lambda$CDM & Lambda Cold Dark Matter\\
{IPAC} & {Infrared Processing and Analysis Center} \\
{LAMBDA} & {Legacy Archive for Microwave Background Data Analysis}\\
{HEASARC} & {High Energy Astrophysics Science Archive Center}\\
{DECaLS} & {Dark Energy Camera Legacy Survey}\\
{BASS} & {Beijing--Arizona Sky Survey}\\
{MzLS} & {Mayall z-band Legacy Survey}\\
{NSF} & {National Science Foundation}\\
{NOIRLab} & {National Optical-Infrared Astronomy Research Laboratory}\\
{LBNL} & {Lawrence Berkeley National Laboratory}\\
{AURA} & {Association of Universities for Research in Astronomy}\\
{DECam} & {Dark Energy Camera}\\
{DES} & {Dark Energy Survey}\\
{ETH} & {Eidgenossische Technische Hochschule}\\
{IEEC} & {Institut d’Estudis Espacials de Catalunya}\\
{CSIC} & {Consejo Superior de Investigaciones Científicas}\\
{SLAC} & {Stanford Linear Accelerator Center}\\
{TAP} & {Telescope Access Program}\\
{NEOWISE} & {Near-Earth Object Wide-field Infrared Survey Explorer}\\
{DOE} & {Department of Energy}\\
{NOAO} & {National Optical Astronomy Observatory}\\
{NAOJ} & {National Astronomical Observatory of Japan}\\
{Kavli IPMU} & {Kavli Institute for the Physics and Mathematics of the Universe}\\
{KEK} & {\textls[-25]{Kō Enerugī Kasokuki Kenkyū Kikō (High Energy Accelerator Research Organization)}}\\
{ASIAA} & {Academia Sinica Institute for Astronomy and Astrophysics}\\
{JSPS} & {Japan Society for the Promotion of Science}\\
{JST} & {Japan Science and Technology Agency}\\
{PS1} & {Pan-STARRS1}\\
{ELTE} & {Eötvös Loránd {Tudom{\'a}nyegyetem} 
}
\end{longtable}
}


\appendixtitles{no} 
\appendixstart
\appendix
\section[\appendixname~\thesection]{}{\label{app}}

\begin{figure}[H]
\begin{adjustwidth}{-1.5cm}{-1.5cm}
\centering
\includegraphics[
    width=1.1\textwidth,
    height=0.939\textheight,
    keepaspectratio
]{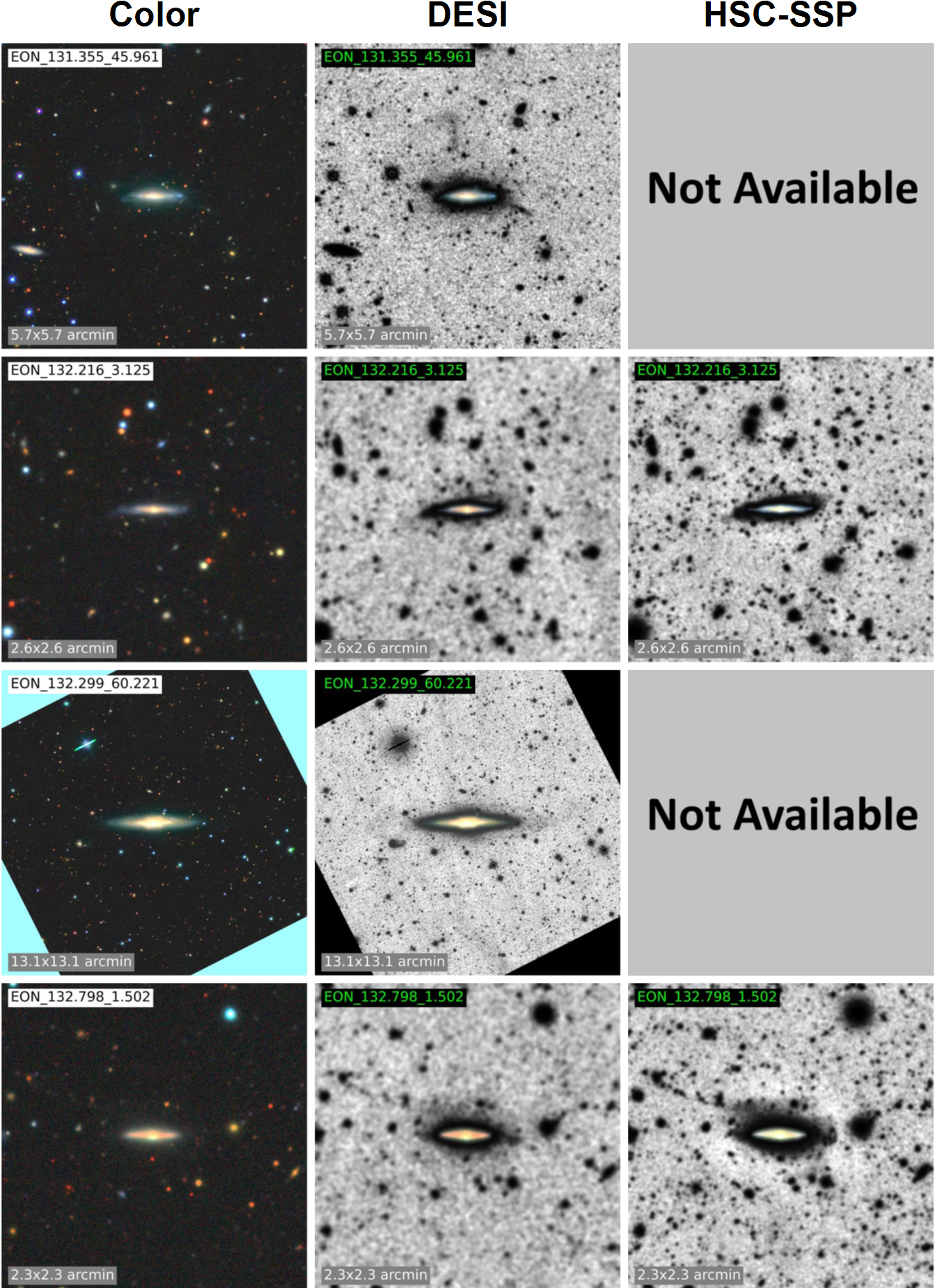}
\end{adjustwidth}
\caption{\textit{Cont}.}
\label{fig:egis_lsb_sample}
\end{figure}

\begin{figure}[H]\ContinuedFloat

\begin{adjustwidth}{-1.5cm}{-1.5cm}
\centering
\includegraphics[width=1.1\textwidth]{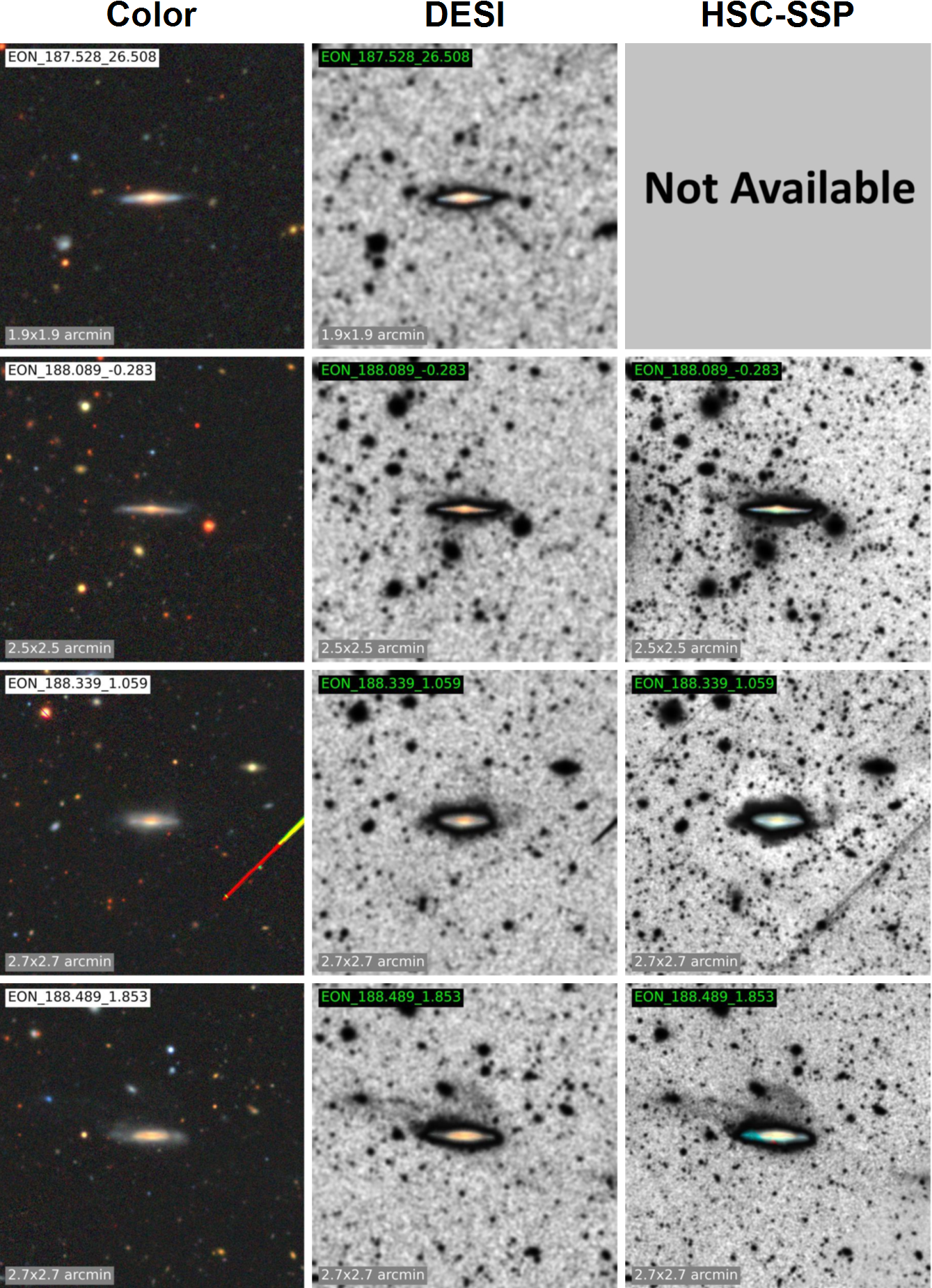}\end{adjustwidth}
\caption{{Sample} from the EGIS catalog of LSB tidal features. Images are presented with DESI, HSC-SSP, and RGB DESI images (created using $g$, $r$, and $z$ bands).}
\end{figure}

\begin{figure}[H]
\begin{adjustwidth}{-1.5cm}{-1.5cm}
\centering
\includegraphics[width=1.1\textwidth]{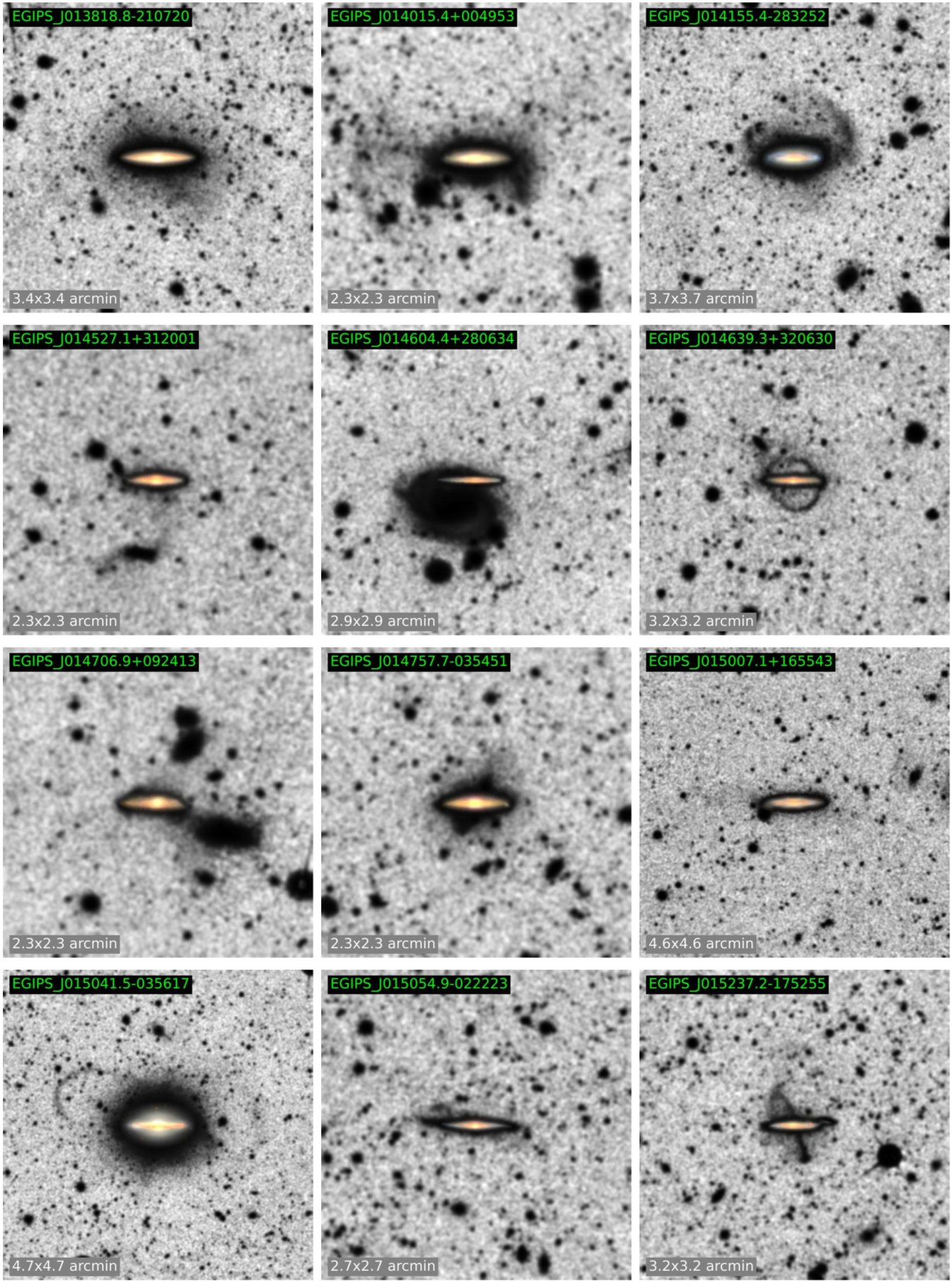}\end{adjustwidth}
\caption{\textit{Cont}.}
\label{fig:egips_lsb_sample}

\end{figure}

\begin{figure}[H]\ContinuedFloat
\begin{adjustwidth}{-1.5cm}{-1.5cm}
\centering
\includegraphics[width=1.1\textwidth]{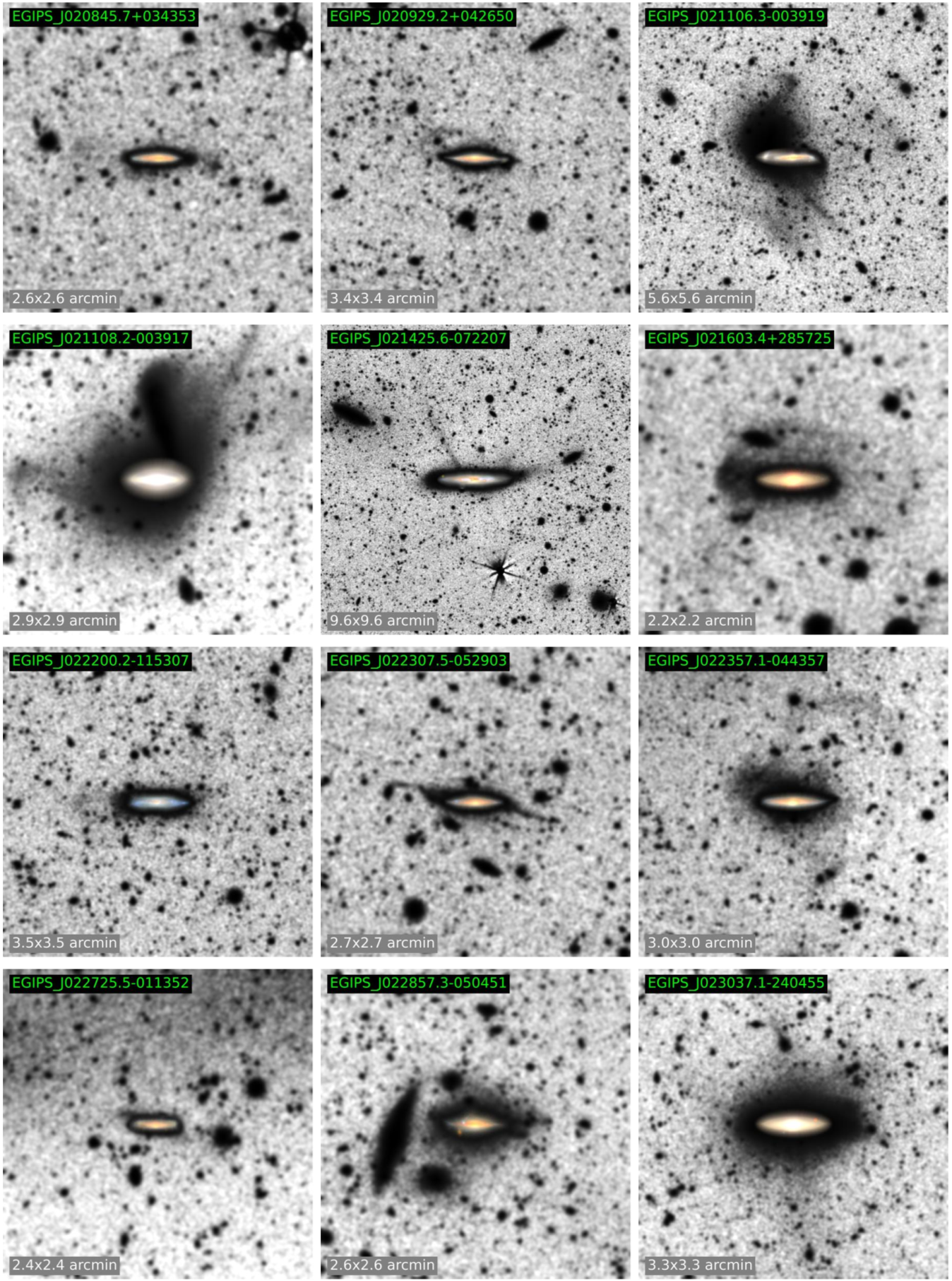}\end{adjustwidth}
\caption{{Sample} from the EGIPS catalog of LSB tidal features. Images are presented with DESI images (created using $g$, $r$, and $z$ bands).}

\end{figure}
\FloatBarrier

\isPreprints{}{

} 

\reftitle{References}

\PublishersNote{}
\isPreprints{}{

} 
\end{document}